\documentclass[sigconf,nonacm]{acmart}
\AtBeginDocument{%
  }

\setcopyright{none}

\begin{document}

\title{JudgmentLens: Human-AI Sensemaking of Complex Legal Judgments} 

\author{Xinyi Chen}
\authornote{Both authors contributed equally to this research.}
\affiliation{%
  \institution{The Hong Kong University of Science and Technology (Guangzhou)}
  \country{China}}
\email{xchen822@connect.hkust-gz.edu.cn}

\author{Ruijie Li}
\authornotemark[1]
\affiliation{%
  \institution{The Hong Kong University of Science and Technology (Guangzhou)}
  \city{Guangzhou}
  \country{China}}
\email{rli541@connect.hkust-gz.edu.cn}

\author{Yuelu Li}
\affiliation{%
  \institution{The Hong Kong University of Science and Technology (Guangzhou)}
  \city{Guangzhou}
  \country{China}}
\email{yli883@connect.hkust-gz.edu.cn}

\author{Chen Liang}
\authornote{Corresponding Author.}
\affiliation{%
  \institution{The Hong Kong University of Science and Technology (Guangzhou)}
  \city{Guangzhou}
  \country{China}}
\email{lliangchenc@163.com}

\begin{abstract}

Judicial judgments are increasingly available, yet dense language and
distributed relationships among facts, evidence, reasoning, and rulings remain
difficult for non-experts to interpret. Through a mixed-methods formative
study with Chinese non-expert readers (survey $N=34$; interviews $N=6$), we
identified structural, interpretive, verification, and action breakdowns. We
developed \textsc{JudgmentLens}, an AI-augmented reading system combining
persistent case representations, adaptive explanations, and traceable links
from generated interpretations to judgment passages. In a counterbalanced
within-subject evaluation ($N=16$), participants completed tasks faster with
\textsc{JudgmentLens} than with conventional PDF reading and reported lower
workload and greater self-reported decision understanding, while rubric-scored
comprehension did not differ reliably. An exploratory PDF+DeepSeek probe suggested that conversational AI supported
formulated questions while leaving question formulation, answer integration,
and source checking largely to users. We contribute an empirical account of
non-expert judgment sensemaking and design strategies for inspectable,
source-grounded AI mediation.

\end{abstract}



\begin{teaserfigure}
  \centering
  \includegraphics[width=\textwidth]{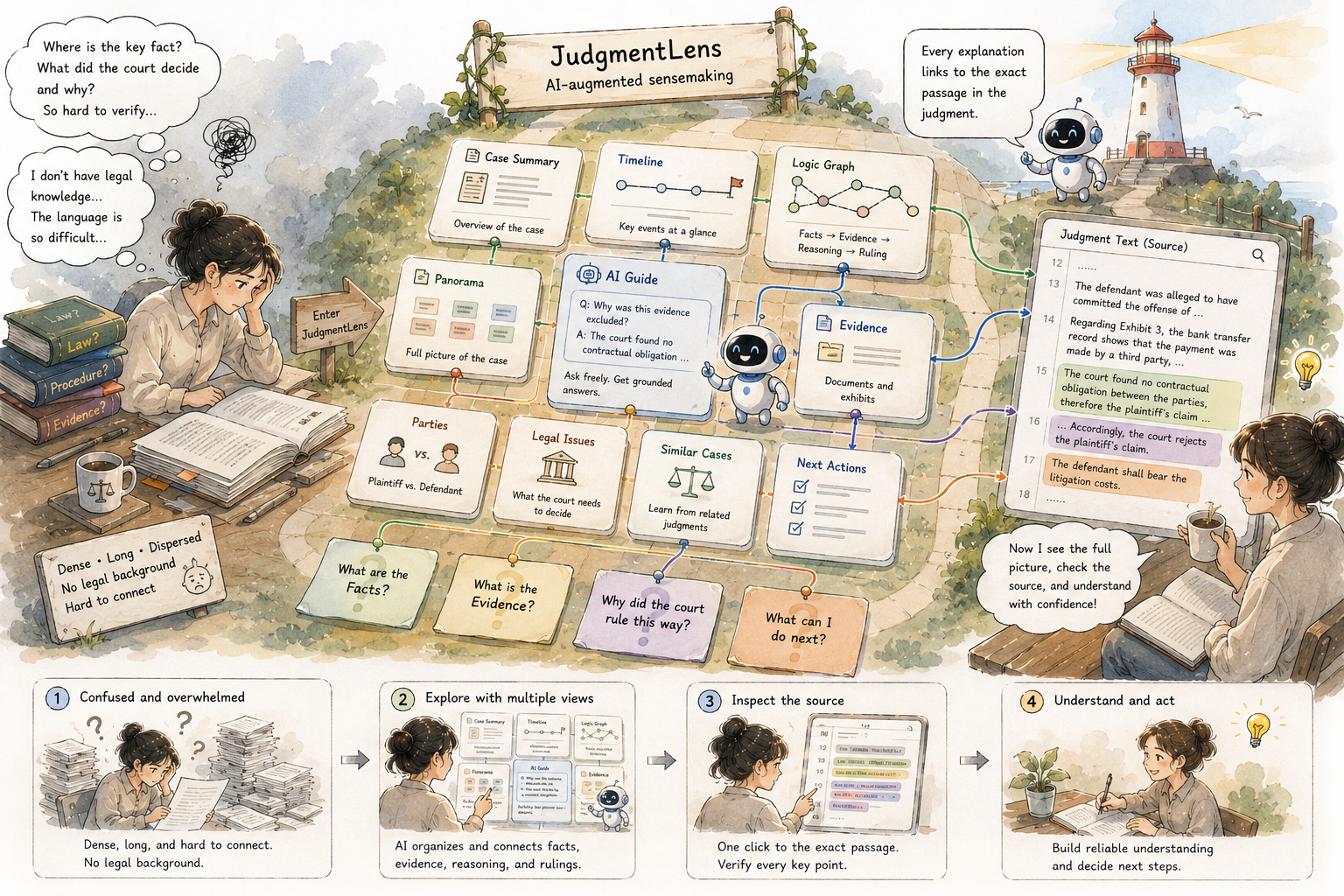}
  \caption{\textsc{JudgmentLens} supports non-expert readers in moving from an overwhelming legal judgment toward structured understanding. The illustrated journey highlights four recurring needs: orienting to the case, connecting evidence and reasoning, verifying interpretations against the source, and preparing for possible next steps.}
  \Description{A watercolor-style conceptual illustration of the reading journey supported by JudgmentLens. On the left, a reader without legal training appears overwhelmed by a long judgment, stacks of documents, and questions about key facts, the court's decision, and verification. A winding path in the center contains connected cards labeled Case Summary, Timeline, Logic Graph, Panorama, AI Guide, Evidence, Parties, Legal Issues, Similar Cases, and Next Actions. Colored paths connect several cards to highlighted passages in a Judgment Text panel on the right, where the reader appears more confident. Four panels along the bottom summarize the transition: confused and overwhelmed; explore with multiple views; inspect the source; and understand and act. Two small robot guides and a lighthouse visualize guidance and source-grounded navigation.}
  \label{fig:teaser}
\end{teaserfigure}

\maketitle

\section{Introduction}
\label{sec:introduction}

The digitization of judicial information has substantially expanded public
access to court judgments in China. Platforms such as China Judgments Online
have made a large volume of judgments searchable and downloadable,
advancing formal judicial transparency ~\cite{weimin2020judicial,liebman2020mass,shi2021smart}. Yet these judgments are primarily disclosed as formal legal documents for institutional legal practice. For people without legal training, access does not imply understanding: specialized terminology, compressed reasoning, and dispersed relationships among facts, evidence, legal rules, and decisions require readers to reconstruct meaning from a structure not designed for non-expert readers. The central
challenge is therefore not simply accessing legal information, but developing a coherent understanding of how and why a court reached a particular decision.

This challenge can be understood as a sensemaking problem~\cite{weick1995sensemaking}. HCI research has
long shown that information-intensive tasks require users to organize
distributed evidence, externalize intermediate representations, and iteratively
refine their understanding rather than merely retrieve relevant information
~\cite{russell1993cost,pirolli1995information,card1999readings,hollan2000distributed}. Accordingly,
prior systems have explored interactive overviews, concept relationships, and
visual analytics to help users navigate and interpret complex documents~\cite{gorg2012combining,liu2014survey,sedig2013interaction,wise1995visualizing,dowling2019interactive}.
For example, ConceptEVA combines automated summarization with interactive
concept visualization, allowing readers to inspect relationships between a
source document and its generated summary ~\cite{zhang2023concepteva}. In the
legal domain, the Legal Macroscope and its Knowlex system use interactive
graphs and maps to help users explore heterogeneous legal sources
~\cite{lettieri2017legal}. Such systems demonstrate the value of externalizing
document structure and relationships. However, they primarily support
document exploration, summarization, or analysis across information
collections. They provide less support for a non-expert reader who is trying
to understand a single authoritative judgment, particularly the relationship
between evidence evaluation, judicial reasoning, and the final decision.

Generative AI further changes the design space by enabling systems to
summarize long documents, answer questions, and generate explanations in
natural language. Recent legal NLP research has demonstrated increasingly
capable models for legal language understanding and legal reasoning
~\cite{chalkidis2022lexglue,guha2023legalbench,hoque2024hallmark}. Human-centered AI research also emphasizes adaptive assistance, mixed-initiative interaction, and the need to preserve human control over AI-supported work~\cite{amershi2019guidelines,horvitz1999principles,shneiderman2022human}. Legal systems such as JusticeBot illustrate how computational assistance can help non-experts explore legal rights, relevant cases, and possible next steps~\cite{westermann2023justicebot}. Yet generative assistance introduces a
different problem: a fluent explanation may be easier to consume while
remaining difficult to verify. When an AI reorganizes a judgment into an
answer, users may have to determine which parts are explicitly supported,
which are synthesized across passages, and whether the explanation faithfully
reflects the authoritative document. Hallucination and unsupported generation
further increase this burden~\cite{miller2019explanation,maynez2020faithfulness,ji2023survey}. Thus,
adding conversational AI to a legal document does not by itself resolve the
sensemaking problem; users still need to understand relationships within the
case and evaluate the basis of AI-generated interpretations.

These observations reveal two gaps in current HCI research. First, although
prior work has studied complex-document sensemaking and has developed legal
information and decision-support systems for non-experts, we still have a
limited empirical understanding of the specific sensemaking breakdowns that
occur when non-experts interpret judgments. Existing studies have
addressed document exploration, legal issue identification, or support for
particular legal tasks, but provide less insight into how non-experts
reconstruct institutional reasoning from the distributed contents of an
individual judgment. Second, explanation research suggests that explanations should respond to users' contextual questions and support iterative inquiry, rather than functioning only as static artifacts attached to model outputs
~\cite{miller2019explanation,liao2020questioning}. For high-stakes authoritative documents, users need not only help generating interpretations, but also mechanisms for inspecting their basis, questioning them, and returning to the original evidence~\cite{liao2020questioning}. These gaps motivate our investigation of how AI can augment judicial judgment interpretation while preserving users' ability to evaluate and control the meaning they construct.

In this work, we investigate the following research questions:

\begin{itemize}
    \item \textbf{RQ1:} What sensemaking barriers do non-experts encounter when
    interpreting complex judicial judgments?
    
    \item \textbf{RQ2:}  How can an AI-augmented judgment-reading system be designed to scaffold non-experts' sensemaking while keeping system-mediated interpretations inspectable against the authoritative source? 
\end{itemize}

To answer these questions, we conducted a mixed-methods formative study with
Chinese non-expert readers who had firsthand experience interpreting judicial
judgments ($N=34$ survey, $N=6$ interviews). We identified four recurring
sensemaking breakdowns: structural breakdown in reconstructing relationships
among facts, evidence, and reasoning; interpretive breakdown in understanding
why decisions were reached; verification breakdown in assessing the
reliability and source basis of AI assistance; and action breakdown in
translating understanding into subsequent steps. These findings motivated
\textsc{JudgmentLens}, an AI-augmented interactive reading system that
combines structured case representations, coordinated visual scaffolding,
adaptive explanations, and traceable connections between AI-generated
interpretations and original judgment passages.  We then conducted a counterbalanced within-subject evaluation with non-expert readers ($N=16$), comparing JudgmentLens with conventional PDF reading. A subsequent exploratory contrastive probe used PDF+DeepSeek to examine how participants experienced question-centered conversational support. Because this probe used a different case and a fixed later position, we interpret it descriptively and qualitatively rather than as a causal baseline comparison. Across the evaluation, we examined task performance, self-reported understanding and workload, and participants' experiences of case organization, source verification, and interpretive control.

This work contributes to HCI in three ways. First, we conceptualize
AI-assisted interpretation of judgments as a human-AI sensemaking
problem, showing why relationship reconstruction and source verification
matter beyond information retrieval or text simplification. Second, we present
\textsc{JudgmentLens} and derive design strategies for AI-augmented
interpretation of authoritative documents, including structural scaffolding,
adaptive explanation, and source-grounded AI mediation that keeps generated
interpretations inspectable against their documentary basis. Third, through formative research and a controlled PDF comparison, complemented by an exploratory conversational-AI probe, we provide quantitative evidence about efficiency, workload, and self-reported understanding, together with qualitative and descriptive insights into how non-experts organize case information, return to source passages, and experience conversational versus persistent structural support.

\section{Related Work}
\label{sec:related_work}

\subsection{Judicial Judgment Interpretation as a Sensemaking Challenge}
\label{subsec:judicial_sensemaking}

Judicial judgments are authoritative documents whose meanings emerge from
relationships among factual narratives, evidence, legal rules, judicial
reasoning, and final decisions. For readers without legal training,
understanding such documents can be difficult because legal texts combine
specialized vocabulary with syntactically and structurally complex forms of
reasoning~\cite{mellinkoff2004language,tiersma1999legal}. Recent empirical work
also suggests that the complexity of legal language can function as part of
how legal documents communicate authority, creating additional comprehension
costs for non-experts~\cite{martinez2024even}. These difficulties are
consequential for access to justice, because people who must interpret legal
documents without professional support may struggle to understand their
rights, obligations, and available courses of action~\cite{sandefur2009access,rhode2004access}.

HCI provides a useful lens for understanding this problem as sensemaking.
Sensemaking research characterizes complex information work as an iterative
process in which people gather distributed evidence, construct intermediate
representations, and progressively refine interpretations
~\cite{weick1995sensemaking,russell1993cost,pirolli2005sensemaking}. In such
tasks, information access alone does not ensure understanding: users must
organize relationships among heterogeneous information elements and maintain a coherent mental model of the information space~\cite{card1999readings,zhang1997nature,hutchins1995cognition,norman2014things}. Judicial judgment interpretation
amplifies this requirement because readers often need to reconstruct how
evidence was treated, how legal rules were applied, and how those intermediate
considerations support a final ruling.

Prior legal technology research has begun to address this challenge through
interactive analysis and decision-support systems. Knowlex, for example,
uses graph and treemap visualizations to help users explore relationships
among legislation, case law, and legal literature~\cite{lettieri2017legal}.
JusticeBot applies a hybrid case-based and rule-based methodology to help
non-experts explore legal rights, relevant cases, and possible next steps in
specific legal situations~\cite{westermann2023justicebot}. These systems show
that legal information can be made more navigable and actionable through
interactive structures. However, they generally target legal information
retrieval, corpus-level exploration, or rule-oriented decision support rather
than the close reading of a single authoritative judgment. They therefore
leave open how non-experts can reconstruct case-specific relationships among
evidence, judicial reasoning, and the final decision within one document.

\subsection{Interactive Support for Complex Document Sensemaking}
\label{subsec:complex_document_sensemaking}

HCI research has developed a broad range of techniques for reducing the
cognitive costs of complex-document understanding. Information visualization
has long explored overviews, filtering, zooming, semantic relationships, and
multiple coordinated views as mechanisms for helping users organize and reason
about large information spaces~\cite{shneiderman2003eyes,card1999readings,heer2012interactive}.
Research on external representations further shows that the structure of a
representation affects what information users can perceive and what
relationships they can readily discover~\cite{zhang1997nature,zhang1995representational,kirsh2009interaction,scaife1996external}. From a distributed-cognition perspective, such representations shift part of the work of maintaining and manipulating relationships from memory to the external environment~\cite{hutchins1995cognition,norman2014things}.

One line of work emphasizes text reduction and summarization. Legal NLP
research has investigated extractive and abstractive summarization as well as
legal language understanding, while studies of plain-language legal
summarization have explicitly targeted non-expert comprehension
~\cite{bhattacharya2019comparative,manor2019plain,urchs2022simplify,billi2025chain,bindal2025lay}.
These approaches can reduce the amount of text that readers must process, but
legal summarization can require substantial abstraction and compression, and
simplification can lose contextual information that matters for interpretation
~\cite{manor2019plain,urchs2022simplify}. In high-stakes documents, this
creates an important design consideration: reducing textual volume is not
equivalent to making the underlying reasoning easier to inspect.

A complementary line of HCI work externalizes document structure through
interactive visual representations. ConceptEVA, for example, combines
automatic summarization with a concept network that allows readers to explore
relationships among concepts and use them to steer generated summaries
~\cite{zhang2023concepteva}. In the legal domain, Knowlex similarly represents
relations among heterogeneous legal sources through interactive graphs and
treemaps~\cite{lettieri2017legal}. Such systems demonstrate how visual
structures can help readers navigate relationships that would otherwise remain
implicit in linear text.

More recent work has begun combining intelligent augmentation with legal
document reading. TermSight, for instance, uses AI-generated plain-language
snippets, visual summaries, and contextual definitions to make legally binding
Terms of Service more approachable, while maintaining links to the original
contract text~\cite{huang2026termsight}. This work is particularly relevant
because it demonstrates the value of placing AI assistance alongside, rather
than instead of, authoritative legal text. At the same time, Terms of Service
reading differs from judicial judgment interpretation: the former centers on
understanding contractual clauses, whereas the latter requires reconstructing
an institutional decision from interconnected facts, evidence, legal rules,
and judicial reasoning. Existing systems therefore provide strong evidence
for structural and contextual augmentation, while leaving open how such
techniques should be organized around the reasoning structure of judicial
decisions for non-expert readers.

Prior work suggests that effective complex-document support benefits from combining textual reduction, externalized structure, and interactive exploration, while many systems remain tied to predefined schemas or relatively stable document structures ~\cite{kang2012examining,zhang2023concepteva,lettieri2017legal,ke2025large}. Generative AI offers more dynamic representations, but also introduces a new interpretive layer whose provenance must remain inspectable.

\subsection{AI-Augmented Sensemaking and Epistemic Agency}
\label{subsec:ai_sensemaking}

Generative AI substantially expands the possibilities for interactive document
support by enabling systems to summarize, explain, answer questions, and adapt
content to users' needs~\cite{kabbara2025ai,hsu2025exploring,aliannejadi2024interactions,ooi2025potential}. In the legal domain, benchmarks such as LexGLUE and
LegalBench demonstrate the increasing capabilities of transformer-based models
and large language models on diverse legal language and reasoning tasks
~\cite{chalkidis2022lexglue,guha2023legalbench}. These capabilities make LLMs
promising mediators for complex legal information, particularly when users need
to ask questions that cannot be anticipated through predefined interfaces.

Human-AI interaction research, however, shows that capability alone does not
determine whether AI assistance will be beneficial~\cite{fan2022human,jiang2021supporting,xu2023transitioning,vaccaro2024combinations,gao2023coaicoder}. AI outputs can be uncertain
or incorrect while remaining persuasive, creating challenges for appropriate
reliance and trust calibration~\cite{parasuraman2000model,bansal2021does,buccinca2021trust}. This concern is especially important in
high-stakes domains, where users may have limited expertise with which to
detect unsupported interpretations. Research on hallucination further
demonstrates that fluent language models may generate statements that are
factually unsupported or inconsistent with their source material~\cite{ji2023survey}.

These concerns have motivated work on explainable and human-centered AI.
XAI research has emphasized explanations as a means of helping users understand
model behavior, while also highlighting the mismatch that can occur between
technical notions of explanation and the questions users actually need answered
~\cite{miller2019explanation,liao2020questioning,kulesza2013too,xu2019explainable,das2020opportunities,adadi2018peeking,chamola2023review,dwivedi2023explainable}. Human-centered AI research similarly advocates designs that preserve human oversight, user control, and appropriate interaction with AI capabilities rather than treating automation
as an end in itself~\cite{amershi2019guidelines,shneiderman2022human}. Provenance research provides a complementary perspective by emphasizing explicit records of where information comes from and how it is transformed~\cite{moreau2022provenance,buneman2001and,herschel2017survey,perez2018systematic,vancisin2023provenance,simmhan2005survey}.

For high-stakes document interpretation, these ideas imply that AI assistance
should remain inspectable against the information from which an interpretation
is derived~\cite{do2024facilitating,moreau2022provenance,buneman2001and}. A generated explanation may combine evidence from multiple
passages, abstract details, or introduce unsupported claims; users therefore
need mechanisms for returning to the underlying source, comparing the
interpretation with its context, and deciding how much reliance is appropriate~\cite{martin2026papertrail}. Recent augmented-reading research points in this direction by foregrounding original legal text alongside AI-generated explanations~\cite{huang2026termsight}. Yet prior work has more often examined either general AI assistance or specific legal document types such as contracts.
Less is known about how source-grounded AI mediation can support the case-level reconstruction of judicial reasoning for readers without legal training.

We characterize this requirement in terms of \emph{epistemic agency}: the
user's ability to inspect information, evaluate the basis of an interpretation,
question AI-generated claims, and retain control over the conclusions they
form~\cite{bennett2023does,shneiderman2022human,amershi2019guidelines}. This framing connects human-AI interaction with document sensemaking:
the goal is not only to provide an answer, but to preserve the user's ability
to understand how an answer relates to authoritative evidence. The design
challenge is consequently to integrate AI-generated interpretation with
interactive structural representations and source-level verification within a
single coherent reading process.

Our work builds on these strands by investigating this integration in judicial
judgment interpretation. We first empirically examine the sensemaking
breakdowns experienced by non-expert readers and then instantiate the resulting
design implications in \textsc{JudgmentLens}, which combines structured case
representations, coordinated visual scaffolding, adaptive explanation, and
source-grounded AI assistance. In doing so, we extend prior work from
information access and document augmentation toward AI-mediated sensemaking in
which generated interpretations remain connected to the authoritative
document and available for human scrutiny.

\section{Formative Study}
\label{sec:formative}

To understand how non-expert readers interpret judgments and identify
opportunities for AI-supported sensemaking, we conducted a two-phase formative
study. Phase~1 used an online questionnaire to map participants' general
judgment-reading practices, perceived difficulties, coping strategies,
emotional experiences, and expectations for AI assistance. Phase~2 followed up
with semi-structured interviews to develop a deeper understanding of how these
difficulties emerged in practice and how participants evaluated possible forms
of support. Thus, the questionnaire provided breadth across the target
population, while the interviews provided depth for interpreting and
synthesizing the resulting design implications.

The study protocol was reviewed and approved by the authors' institutional
review board prior to data collection. All participants provided informed
consent and were informed of their right to withdraw. Because judicial
judgments may contain sensitive personal information, we did not collect
participants' original judgments or personally identifiable legal materials.
Interview recordings and transcripts were anonymized before analysis.

\subsection{Phase 1: Online Questionnaire Survey}
\label{subsec:phase1_questionnaire}

\subsubsection{Participants}

We recruited participants through online community postings and social media
channels. Eligibility required participants to (1) have firsthand experience
participating in litigation or related legal proceedings, (2) have encountered
or interpreted a judicial judgment, and (3) have no formal legal education,
professional legal training, or employment in the legal profession. These
criteria were intended to capture readers who had a practical need to
understand judicial decisions but lacked the domain knowledge and interpretive
strategies typically acquired through legal training.

We collected 34 valid questionnaire responses. Participants reported diverse litigation experiences, including civil (55.88\%), criminal (20.59\%), labor (17.65\%), administrative (2.94\%), and other disputes (2.94\%). In terms of litigation outcomes, 15 participants (44.12\%) reported being on the losing side of their most recent case. Most participants had encountered relatively lengthy judgments: 29 (85.29\%) reported documents exceeding five pages.

\subsubsection{Design and Results}

The questionnaire contained 15 items covering demographic background,
litigation experience, judgment-reading behavior, perceived difficulty,
coping strategies, emotional experience, and expectations for AI-assisted
interpretation. The complete questionnaire
wording and response options are provided in
Appendix~\ref{app:formative-materials}. The questionnaire was primarily descriptive and was used to
identify recurring patterns that would inform the subsequent interviews rather
than to test hypotheses.

Half of the participants (50.00\%, $n=17$) rated the difficulty of
understanding their judgment at 4 or above on a 5-point scale. The most
frequently reported difficult sections were evidence admissibility
(70.59\%, $n=24$), statutory references (67.65\%, $n=23$), and judicial
reasoning (64.71\%, $n=22$). Participants also reported substantial reliance
on external coping strategies, including consulting lawyers or knowledgeable
peers (76.47\%, $n=26$), conducting online searches (67.65\%, $n=23$), and
looking up legal provisions (67.65\%, $n=23$). In addition, 29 participants
(85.29\%) reported experiencing anxiety, helplessness, or dissatisfaction
because of difficulty understanding judgments. At the same time, 31
participants (91.17\%) expressed willingness to use an AI-assisted visual
reading tool. Table~\ref{tab:formative_summary} summarizes the
design-relevant findings.

\begin{table*}[t]
\centering
\caption{Key questionnaire findings related to judicial judgment interpretation
($N=34$).}
\label{tab:formative_summary}
\small
\begin{tabular}{lll}
\toprule
\textbf{Dimension} & \textbf{Measure} & \textbf{Result}\\
\midrule
\textbf{Reading difficulty}
& High perceived difficulty ($\geq4$/5)
& 50.00\% ($n=17$)\\
& Evidence admissibility difficult
& 70.59\% ($n=24$)\\
& Statutory references difficult
& 67.65\% ($n=23$)\\
& Judicial reasoning difficult
& 64.71\% ($n=22$)\\
\midrule
\textbf{Coping strategies}
& Consulted lawyers/knowledgeable peers
& 76.47\% ($n=26$)\\
& Searched online resources
& 67.65\% ($n=23$)\\
& Looked up legal provisions
& 67.65\% ($n=23$)\\
\midrule
\textbf{Emotional experience}
& Anxiety/helplessness/dissatisfaction
& 85.29\% ($n=29$)\\
\midrule
\textbf{AI assistance}
& Willing to use AI visual reader
& 91.17\% ($n=31$)\\
\bottomrule
\end{tabular}
\end{table*}

\subsection{Phase 2: Semi-structured Interviews}
\label{subsec:phase2_interviews}

The second phase was designed to complement the questionnaire by examining
participants' experiences in greater depth. After completing the questionnaire,
respondents were asked whether they were willing to participate in a follow-up
interview. We contacted respondents who expressed interest and purposively
selected six participants to capture variation in litigation context,
outcome, and reading experience. Thus, the interview sample was drawn directly
from the Phase~1 cohort rather than recruited independently. Table~
\ref{tab:interview_participants} summarizes their backgrounds.

\begin{table}[t]
\centering
\caption{Interview participant profiles ($N=6$).}
\label{tab:interview_participants}
\small
\begin{tabular}{cclll}
\toprule
\textbf{ID} & \textbf{Gender} & \textbf{Age} & \textbf{Case Type} &
\textbf{Outcome}\\
\midrule
F1 & Female & 26--35 & Civil (Divorce \& Custody) & Lost\\
F2 & Male & 18--25 & Civil (Fraud / Debt Recovery) & Lost\\
F3 & Female & 18--25 & Civil / Labor (Unpaid Wage) & Won\\
F4 & Male & 36--45 & Civil (Marital Debt Dispute) & Lost\\
F5 & Male & 26--35 & Criminal (Gambling-related) & Lost\\
F6 & Male & 26--35 & Administrative (Police Dispute) & Won\\
\bottomrule
\end{tabular}
\end{table}

Each interview lasted approximately 20--30 minutes and was conducted through
video conferencing. The semi-structured protocol explored how participants
approached a judgment, where interpretation broke down, what strategies they
used when encountering difficulties, what forms of AI assistance they had
previously tried, and what support they expected from an AI-supported reading
system. This phase allowed us to interpret the questionnaire patterns in the
context of participants' reported reading practices and experiences.

All interviews were recorded, transcribed, and anonymized. Two researchers
analyzed the transcripts using thematic analysis following Braun and Clarke
~\cite{braun2006using}. We first performed open coding of interview segments,
then iteratively grouped related codes and refined them into higher-level
themes through discussion. Questionnaire patterns were considered alongside
the interview themes when synthesizing the final formative results. The interview core prompts and a condensed analytical framework
are provided in Appendix~\ref{app:formative-materials}. 

\subsection{Formative Study Results}
\label{subsec:formative_results}

\begin{figure*}[t]
    \centering
    \includegraphics[width=0.95\linewidth]{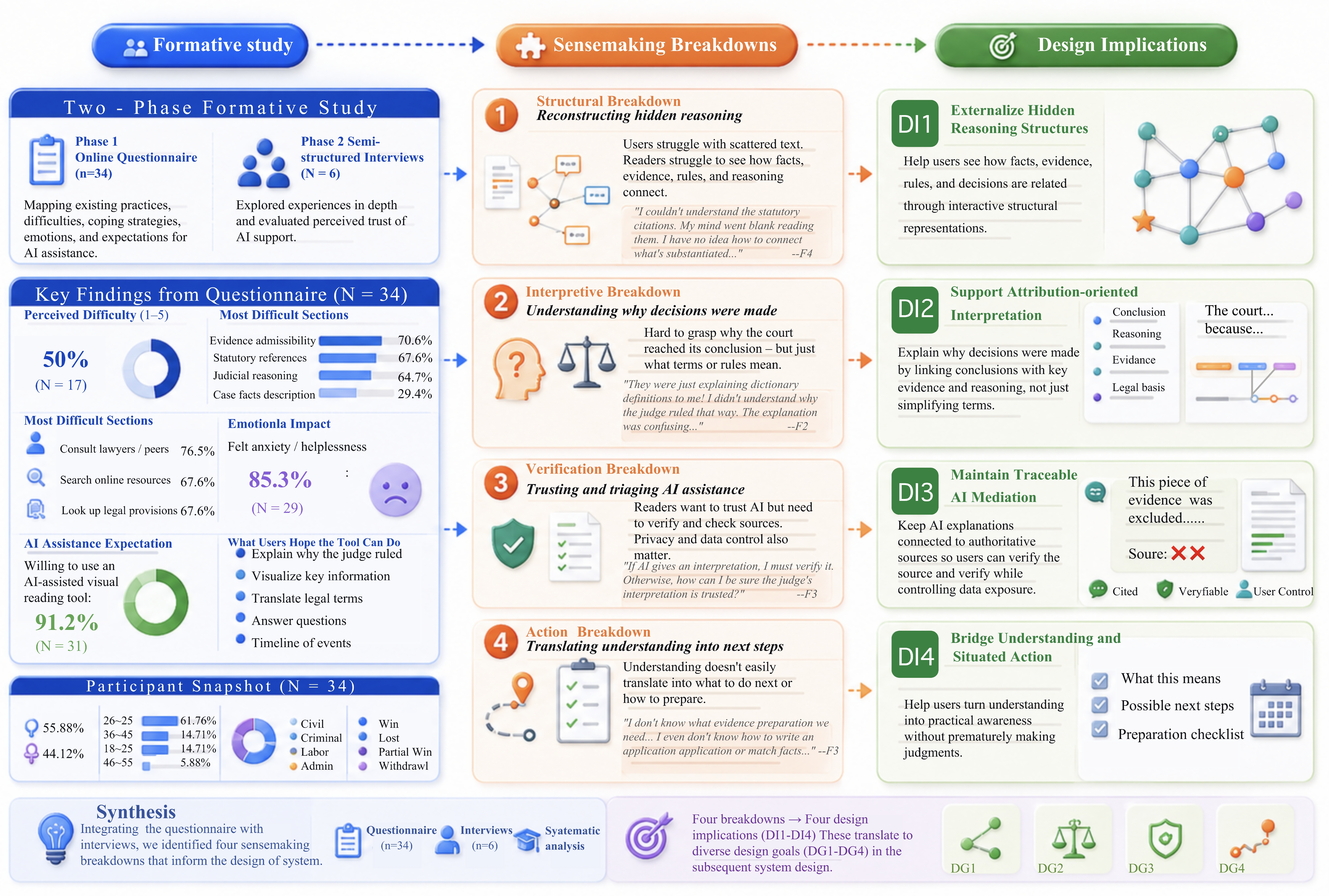}
    \caption{Empirical framework connecting formative-study evidence,
    sensemaking breakdowns, and design implications for AI-augmented judicial
    judgment interpretation. The four breakdowns---structural, interpretive,
    verification, and action---motivate DI1--DI4, respectively, which guide
    the design of JudgmentLens.}
    \label{fig:formative_framework}
\end{figure*}

Synthesizing the broad patterns from Phase~1 with the more detailed accounts
from Phase~2, we identified four recurring sensemaking breakdowns:
\emph{structural breakdown}, where readers struggled to reconstruct
relationships among facts, evidence, and reasoning; \emph{interpretive
breakdown}, where readers struggled to understand why decisions were reached;
\emph{verification breakdown}, where readers questioned whether AI-generated
interpretations were reliable and traceable to the source; and \emph{action
breakdown}, where understanding did not readily translate into subsequent
procedural steps. Figure~\ref{fig:formative_framework} summarizes how these
empirical patterns were synthesized into the design implications that guided
JudgmentLens.

\subsubsection{Structural Breakdown: Reconstructing Hidden Reasoning}

The questionnaire identified evidence admissibility, statutory references, and
judicial reasoning as the most difficult sections. Interview participants
(F1, F4, F5) similarly described difficulty connecting information distributed
across factual narratives, evidence descriptions, and the court's reasoning.
Their accounts suggested that the primary burden was not simply the volume of
information, but the work of reconstructing relationships that remained
implicit in linear text.

F4 captured this difficulty succinctly:
\begin{quote}
\textit{``I couldn't understand the statutory citations. My mind went blank
reading the dense text without a clear line connecting who submitted what.''}
\end{quote}

Together, these findings indicate that non-expert readers often had access to
the relevant information but lacked an explicit representation of the
relationships connecting evidence, claims, and judicial reasoning.

\subsubsection{Interpretive Breakdown: Understanding Why Decisions Were Made}

Beyond locating information, participants needed to understand why the court
reached a particular conclusion. This difficulty was particularly salient
among participants who had experienced unfavorable outcomes (F1, F2, F4, F5).
Their accounts focused less on the definitions of individual legal terms and
more on understanding how the court evaluated their evidence and claims.

F2 described his experience with general-purpose AI assistance:
\begin{quote}
\textit{``They were just explaining dictionary terms... I didn't want a legal
lecture---I wanted to know why the judge ruled that my transfer screenshots
were insufficient.''}
\end{quote}

This pattern suggests that accessible judgment interpretation requires
connecting outcomes with the evidence and reasoning that produced them, rather
than only simplifying legal terminology.

\subsubsection{Verification Breakdown: Trusting and Tracing AI Assistance}

Participants frequently sought external help when encountering incomprehensible
legal content. At the same time, interview accounts revealed an important
boundary around AI assistance (F1, F2, F3, F5): participants were interested
in using AI but remained concerned about whether generated explanations were
faithful to the judgment and whether sensitive documents could be safely
processed.

F3 articulated the need for source traceability directly:
\begin{quote}
\textit{``If AI gives an interpretation, it must pinpoint the exact line in
the original judgment text.''}
\end{quote}

The issue was therefore not simply whether users trusted AI in general, but
whether they could inspect the basis of a particular interpretation. Privacy
concerns further reinforced the need for control over which representation of
a judgment was exposed to AI processing.

\subsubsection{Action Breakdown: Translating Understanding into Next Steps}

Participants viewed judgment interpretation as connected to subsequent
procedural decisions rather than as an isolated reading activity (F3, F6).
They described difficulties understanding what should be prepared after a
judgment, particularly when they lacked professional guidance.

F3 described this difficulty as follows:
\begin{quote}
\textit{``When I started labor arbitration pro se, I didn't even know how to
write an arbitration application or match facts to statutes.''}
\end{quote}

These accounts suggest that users sought assistance that could connect
understanding of the judgment with concrete, reviewable preparation tasks,
rather than replacing human decision making with automated outcome
predictions.

\subsection{Design Implications}
\label{subsec:design_implications}

The four breakdowns were translated into four design implications that directly
guided the subsequent system design:

\begin{itemize}

\item \textbf{DI1: Externalize Hidden Reasoning Structures.}
Systems should help users reconstruct relationships among facts, evidence,
legal rules, and decisions through interactive structural representations.

\item \textbf{DI2: Support Attribution-oriented Interpretation.}
AI assistance should explain why decisions were reached by connecting
conclusions with relevant evidence and reasoning rather than merely simplifying
legal terminology.

\item \textbf{DI3: Maintain Traceable AI Mediation.}
AI-generated interpretations should remain connected to authoritative document
sources, enabling users to inspect, question, and verify generated
explanations while retaining appropriate control over the data exposed to AI.

\item \textbf{DI4: Bridge Understanding and Situated Action.}
Systems should help users translate understood judgments into procedural
awareness and practical preparation without presenting AI-generated outcomes as
substitutes for human or professional legal judgment.

\end{itemize}

These implications directly guide the system design below.

\section{The JudgmentLens System}
\label{sec:system}

Guided by DI1--DI4, we developed \textsc{JudgmentLens}, an AI-augmented workspace that helps non-experts read judgments as structured, inspectable accounts. The system is not designed to replace the judgment with a single summary or to turn judgment reading into a sequence of chatbot answers. Instead, it maintains a set of persistent representations through which readers can establish an overview, examine how evidence and legal considerations relate to the ruling, formulate follow-up questions, and return to the court's text whenever an interpretation requires inspection.

Figure~\ref{fig:teaser} presents a conceptual overview of the sensemaking
process supported by \textsc{JudgmentLens}. It follows a non-expert reader from initial difficulty with a dense judgment, through exploration of coordinated case representations and case-specific AI assistance, to inspection of highlighted source passages and preparation for possible next steps. The central path organizes information around four recurring questions---what the facts are, what evidence matters, why the court ruled as it did, and what the reader may need to do next---while the colored links to the Judgment Text represent the source-tracing interaction shared across the system. In the implemented workspace, these functions are accessed through nine persistent views grouped into \emph{Case Understanding}, \emph{Analysis and Actions}, and \emph{Source Material}; explanation depth, reading perspective, and original or masked text can be adjusted without losing the active case or source context. Moreover, the four stages shown in the teaser summarize an overall transition rather than prescribe a fixed workflow. In practice, readers may move between overview, detail, AI assistance, source inspection, and preparation as needed.

\subsection{Entering and Configuring the Reading Workspace}
\label{subsec:workspace_entry}

Users begin by uploading a judgment in PDF or Word format. Before the workspace is prepared, the entry screen asks how the judgment should initially be presented. \emph{Explanation Depth} provides beginner, standard, and professional modes. Beginner mode prioritizes plain-language orientation and explains legal terms when needed; standard mode retains more of the judgment's terminology and reasoning detail; and professional mode presents denser legal and evidential structure. \emph{Reading Perspective} allows readers to begin with a global view or emphasize the prosecution or defense position. \emph{Text Display} selects either the original extracted text or a version in which common identifying fields are replaced with masked aliases.

These settings alter presentation rather than the underlying case representation. Readers can change them at any time, while the active case and source links are preserved; the original judgment remains directly accessible.

Preparation runs asynchronously after upload. The interface can expose completed views progressively, while the Judgment Text remains available if a generated module is delayed or unavailable. Returning users can reopen a prepared case and continue from their previous reading configuration. These behaviors make document preparation part of the reading workspace rather than a blocking, one-shot generation step.

\subsection{Constructing a Case-level Mental Model}
\label{subsec:case_model}

\begin{figure*}[t]
  \centering
  \includegraphics[width=\textwidth]{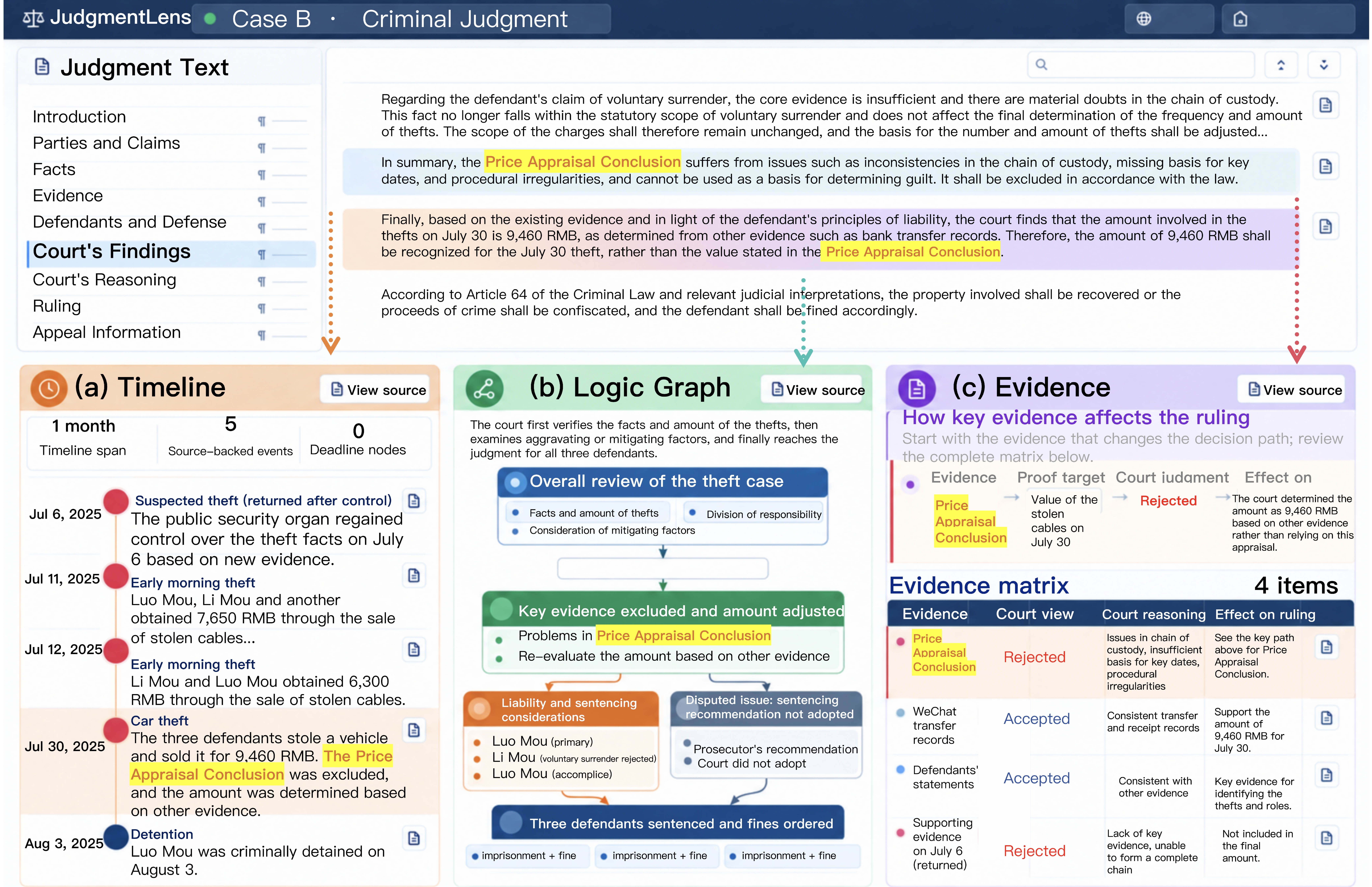}
  \caption{Coordinated representations for constructing a case-level mental model: (a) the Timeline orders source-backed factual and procedural events; (b) the Logic Graph connects facts and evidence to judicial considerations, disputed issues, and the ruling; and (c) the Evidence view separates an item's proof target, judicial evaluation, stated reasoning, and effect on the ruling. Source-linked controls provide a contextual path back to the judgment.}
  \Description{Three JudgmentLens interface views. The timeline lists five dated events. The logic graph connects case questions, facts and evidence, sentencing considerations, a disputed issue, and the final ruling. The evidence matrix compares four evidence items across their proof targets, court evaluations, reasoning, and effects on the ruling.}
  \label{fig:structured-views}
\end{figure*}

The Case Understanding region contains five coordinated views that address different questions a non-expert may ask while reconstructing a judgment: \emph{What is this case about? What happened and when? How did the court connect evidence and rules to its conclusions? What did each item of evidence establish?} Figure~\ref{fig:structured-views} illustrates the three most structurally distinctive views.

The \emph{Case Summary} provides an entry point rather than a replacement judgment. It groups the operative outcome with the person to whom it applies, presents a one-sentence case thread, and lists the court's principal stated reasons. The organization is especially useful in multi-party cases, where sentences, monetary obligations, and findings can otherwise be confused across defendants. Counts and expansion controls preserve access to additional parties, reasons, or rulings instead of silently omitting them from the compact view.

The \emph{Timeline} reconstructs factual and procedural events in chronological order. Each event records what occurred, why it matters to the case, and whether it affects a charge, evidential issue, deadline, or outcome. In the example shown in Figure~\ref{fig:structured-views}a, the timeline distinguishes an accusation that was later withdrawn, two theft events, a subsequent disputed valuation event, and the defendants' detention. This view makes temporal dependencies persistent rather than requiring readers to repeatedly search across the narrative.

The \emph{Logic Graph} provides a complementary inferential view. It connects case questions to relevant facts and evidence, judicial considerations, disputed issues, and the operative ruling. For example, Figure~\ref{fig:structured-views}b shows how the court first established the theft events and amount, excluded a procedurally defective price-appraisal conclusion, used other evidence to determine the amount, considered differences in the defendants' roles and mitigating circumstances, and arrived at individual sentences. The graph is descriptive: it externalizes relationships expressed in the judgment but does not propose an alternative ruling or assert what the court should have decided.

The \emph{Evidence} view focuses on a distinction that participants in the formative study often found difficult: evidence being submitted does not imply that the court accepted it or relied on it. The matrix therefore separates the evidence item, its proof target, the court's acceptance or rejection, the court's stated reason, and its effect on the ruling. A compact path above the matrix foregrounds evidence that materially changes the decision path. In Figure~\ref{fig:structured-views}c, the price-appraisal conclusion is marked as rejected because of procedural and authenticity concerns, while transfer records and defendant statements are shown as accepted for different purposes.

Finally, the \emph{Case Panorama} brings facts and charges, key evidence, judicial views, and the operative ruling into a single board. Whereas the Summary prioritizes orientation, the Timeline prioritizes temporal order, the Logic Graph prioritizes inferential relationships, and the Evidence view prioritizes evidential status, the Panorama provides a compact cross-sectional account. These are coordinated projections of one case representation. Readers can move among them according to the question at hand while retaining the same source-linked entities and case context (DI1).

\subsection{Moving between Representations, Questions, and Sources}
\label{subsec:grounded_interaction}

All primary representations use the same source-linking interaction. A document icon appears on reviewable summary cards, timeline events, graph nodes, panorama items, evidence rows, and preparation items. Selecting it updates the workspace's source focus, opens the Judgment Text panel on the right, scrolls to the corresponding paragraph, and highlights the most directly related text. The surrounding paragraphs remain visible so that a reader can inspect whether the selected excerpt has been interpreted out of context. Closing the panel returns space to the active representation without resetting the selected depth, perspective, scroll position, or graph viewport.

This mechanism follows DI2 and DI3 by reducing the navigation cost of verification rather than claiming that a generated representation is necessarily correct. Readers can also open the Judgment Text independently, browse its section outline, search within the document, enlarge the text, and enable synchronized highlighting. Thus, source access is not conditional on accepting an AI-generated view: contextual links provide a shortcut from a specific interpretation, while direct access allows readers to bypass mediation altogether.

The \emph{AI Guide} applies the same source-grounding principle to free-form
questions about the active case. Because it operates within the active case, users do not need to re-upload the judgment or reconstruct its context in a separate chatbot. The interface offers case-specific question starters, streams the response, and links its principal claims to corresponding judgment passages. A collapsible \emph{How AI Reached this Answer} panel describes the source-use steps, and \emph{Check source} opens the same contextual source panel used elsewhere in the workspace. The interface summarizes the
categorical source-support status under the label \emph{AI Confidence}. This label is not a confidence score, a calibrated probability, or an estimate of factual correctness~\cite{bansal2021does,buccinca2021trust}. When source support is incomplete, the interface prompts further review rather than treating fluent output as authoritative. Suggested follow-up questions allow users to refine an inquiry while the persistent Summary, Timeline, Logic Graph, and Evidence views remain available around the conversation.

\begin{figure*}[t]
  \centering
  \includegraphics[width=\linewidth]{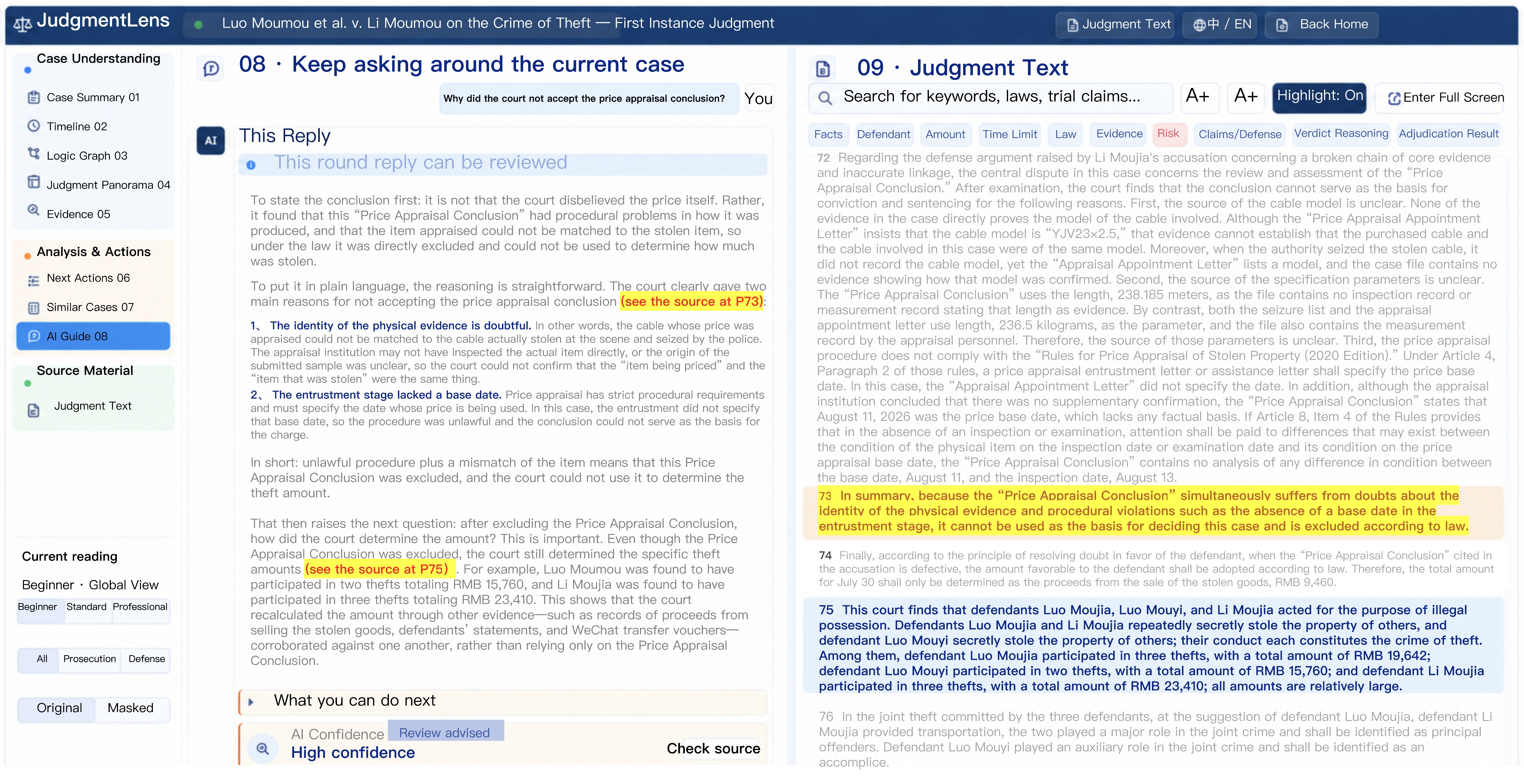}
  \caption{Source-grounded follow-up in the AI Guide. A response retains case-specific source references and a \emph{Check source} action. Following a reference opens the Judgment Text at the corresponding paragraph while preserving the reader's current case representation and settings.}
  \Description{Two connected interface details. The first shows an AI Guide response with source references, a review cue, and a Check source button. The second shows the source panel positioned at and highlighting the corresponding paragraph in the judgment.}
  \label{fig:ai-source}
\end{figure*}

This creates a different interaction structure from an isolated question--answer interface. Users can formulate questions from already visible case relationships, inspect a response against the source, and return to a persistent representation to integrate the answer into the case-level account. The intended sequence is therefore not simply question~$\rightarrow$~answer, but representation~$\rightarrow$~question~$\rightarrow$~source inspection~$\rightarrow$~continued sensemaking. Figure~\ref{fig:ai-source} illustrates this source-grounded interaction.

\subsection{Extending Understanding toward Preparation}
\label{subsec:preparation}

The Analysis and Actions region explores how an established understanding of the judgment can support subsequent preparation (DI4). \emph{Next Actions} surfaces source-linked deadlines, monetary or procedural obligations, and issues in the judgment that may require checking. Items are organized by temporal urgency, such as an appeal deadline or an obligation that should be verified against the operative text. They are presented as information to review, not as predictions or recommendations about which legal action a user should take.

\emph{Similar Cases} supports bounded comparison. It first states which dimensions of the active judgment may be relevant to comparison and then presents candidate rules or cases with their provenance, points of similarity, points of difference, and limitations. The interface explicitly distinguishes a guiding rule case from a factually similar case and avoids transferring an outcome from one case to another. This view is intended to help readers formulate further questions or prepare for consultation; it does not estimate a probability of success or substitute for professional legal analysis.

Together, these views extend the workspace beyond initial comprehension while preserving the same interaction boundary as the rest of the system: source-backed information can be organized and compared, but the significance of that information and decisions about subsequent action remain with readers and, where appropriate, legal professionals.

\subsection{Implementation of the Source-linked Workspace}
\label{subsec:implementation}

\textsc{JudgmentLens} is implemented as a React/TypeScript web application with a FastAPI backend and asynchronous preparation jobs. After upload, parsers extract judgment text, paragraph boundaries, and case metadata into aligned original and redacted representations. The backend then constructs a versioned \emph{Case Fact Core} containing parties, facts, evidence, reasoning, rulings, and obligations. Summary, timeline, graph, panorama, evidence, and preparation modules are generated as coordinated projections of this shared representation.

Reviewable items retain source anchors containing paragraph identifiers and
supporting excerpts. Selecting an anchor opens and highlights the corresponding context without discarding the active view. The AI Guide similarly retrieves paragraph-level passages before generation~\cite{lewis2020retrieval,gao2023retrieval} and exposes qualitative source-support cues. These cues facilitate inspection but are not calibrated probabilities or guarantees of legal correctness.

Before evaluation, we froze the prompts, schemas, and model routing and
pre-generated the configurations used with Cases A and B. Researchers manually checked participant-facing facts, roles, disputed issues, rulings, obligations, and linked passages against the judgments. This process stabilized the study materials but does not establish general model correctness. Architecture, routing, validation, and preparation details appear in Appendix~\ref{app:implementation}.

\paragraph{Artifact availability.}
The implementation of \textsc{JudgmentLens}, including source code, deployment instructions, prompt and schema versions, and supporting configuration files, will be released through an anonymized project repository for review and made publicly available upon publication.

\section{User Evaluation}
\label{sec:user-evaluation}

\begin{figure*}[t]
    \centering
    \includegraphics[width=\linewidth]{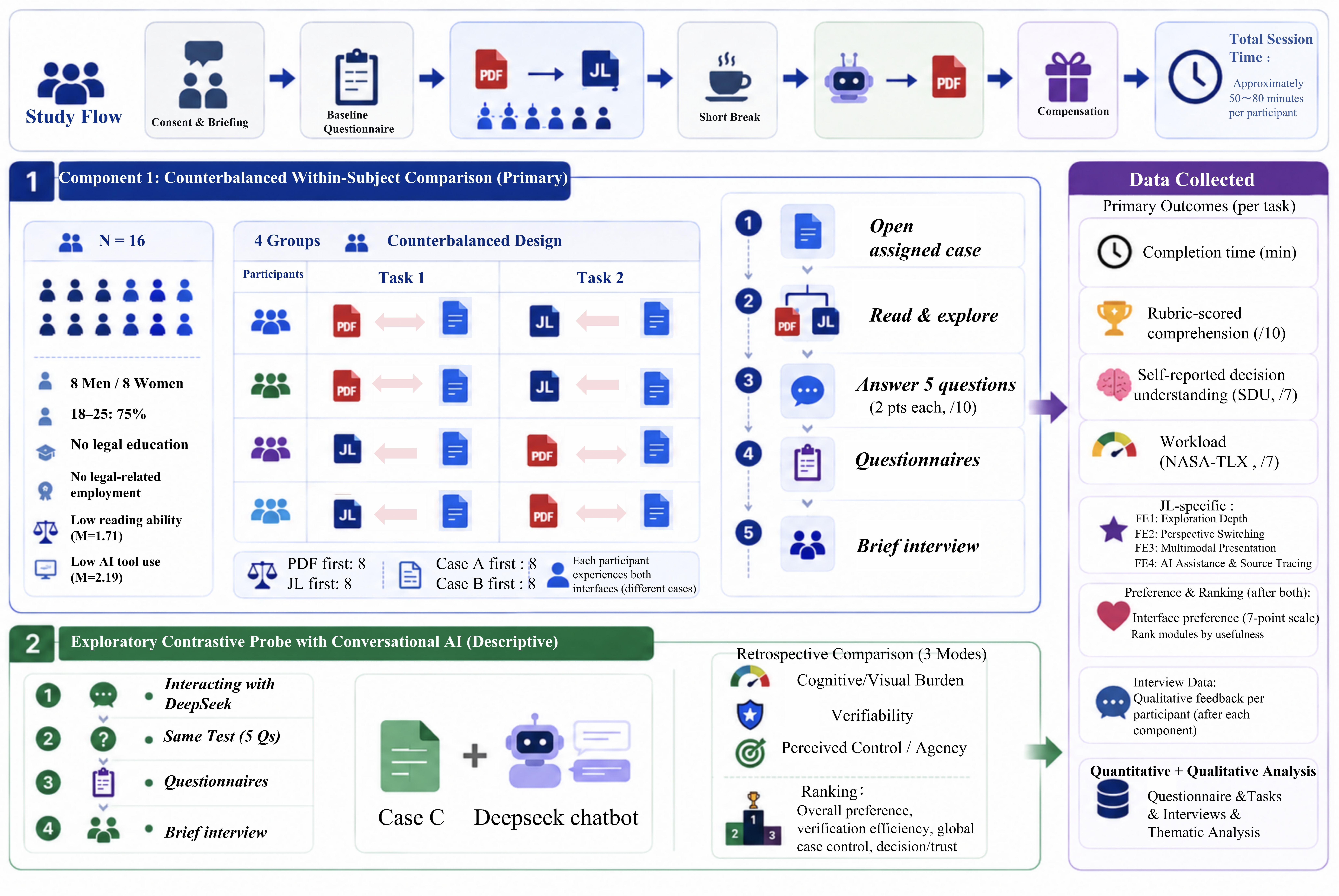}
    \caption{Evaluation procedure. The primary component was a counterbalanced within-subject comparison of PDF and \textsc{JudgmentLens} using Cases A and B. The subsequent PDF+DeepSeek component used Case C as an exploratory contrastive probe; it was not treated as a randomized third condition.}
    \label{fig:evaluation-procedure}
\end{figure*}

We evaluated \textsc{JudgmentLens} through two components with deliberately different evidentiary roles. First, a counterbalanced within-subject comparison examined how \textsc{JudgmentLens} affected reading efficiency, rubric-scored comprehension, self-reported decision understanding, and workload relative to conventional PDF reading. This primary comparison provides the controlled quantitative evidence in this section. Second, we conducted an exploratory contrastive probe with PDF+DeepSeek to examine what sensemaking work remained when participants relied on a general-purpose conversational interface. Because the probe used a different case and was always administered after the controlled comparison, we use it descriptively and qualitatively rather than to estimate a causal interface effect. Figure~\ref{fig:evaluation-procedure} summarizes the two evaluation components and their evidentiary roles.

\subsection{Evaluation Overview and Participants}

The evaluation followed the institutional ethics protocol described in Section~\ref{sec:formative}. We recruited 16 participants (P1--P16) through social media; all provided informed consent and reported no formal legal education or legal-related employment. Eight participants identified as men and eight as women. Twelve (75.0\%) were aged 18--25 and four (25.0\%) were aged 26--35. Ten participants (62.5\%) reported college- or undergraduate-level education, and six (37.5\%) reported master's-level education or above. Eleven participants (68.75\%) reported almost never reading lengthy legal or other complex documents, and five (31.25\%) did so occasionally. On five-point scales, participants reported limited ability to read legal judgments ($M=1.71$) and limited prior use of AI-assisted reading or legal-retrieval tools ($M=2.19$).

The study was conducted remotely through Tencent Meeting. Participants used
their own computers and shared their screens throughout the session; with
their consent, the complete sessions were recorded. Each participant received
RMB~30 upon completion. All participants completed both evaluation components. We recorded and transcribed the brief semi-structured interviews conducted after each component. We used thematic analysis ~\cite{braun2006using} to examine the two sets of transcripts separately before comparing recurring patterns concerning case organization, source verification, conversational-AI use, and interaction friction. Quotations below were translated from Chinese and lightly edited only to remove disfluencies.

\subsection{Controlled Comparison: PDF vs. JudgmentLens}

\subsubsection{Design and Procedure}

Participants completed one task with a conventional PDF reader and one with
JudgmentLens. Cases A and B were both first-instance criminal judgments
concerning jointly committed theft. Each contained approximately 14,000
Chinese characters and occupied 14 PDF pages. We selected them to be
comparable in offense type, document length, and overall structure. We crossed
case and interface order across four groups:

\begin{itemize}
    \item P1, P2, P3, and P13: Case B--PDF $\rightarrow$ Case A--\textsc{JudgmentLens};
    \item P4, P5, P6, and P14: Case A--PDF $\rightarrow$ Case B--\textsc{JudgmentLens};
    \item P7, P8, P9, and P15: Case B--\textsc{JudgmentLens} $\rightarrow$ Case A--PDF;
    \item P10, P11, P12, and P16: Case A--\textsc{JudgmentLens} $\rightarrow$ Case B--PDF.
\end{itemize}

For each task, participants used the assigned interface to inspect the
judgment and answer five comprehension questions covering the case outcome,
disputed issues, evidence interpretation, procedural or temporal obligations,
and subsequent actions. Using a predefined answer key, the first author
assigned 0 points to an incorrect or unsupported answer, 1 point to a
partially correct answer, and 2 points to a complete answer, yielding a maximum
score of 10. Completion time was measured from opening the assigned judgment
and receiving the questions to submitting all five answers. The complete questions, case-specific answer keys, item-level scoring rules, and assessed dimensions are provided in Appendix~\ref{app:evaluation-materials}.

After each task, participants completed the same five-item self-reported decision understanding (SDU) questionnaire on seven-point scales (1=strongly disagree, 7=strongly agree). The items assessed information extraction, logic mapping, evidence association, subsequent inference, and confidence in one's answers; their mean formed the SDU composite. Participants also completed a six-item adaptation of NASA-TLX ~\cite{hart1988development}. We averaged mental demand, temporal demand, effort, frustration, and visual/interaction strain to form a workload composite, with higher values indicating greater burden. The perceived-performance item used the opposite direction (1=very high performance, 7=very low performance) and was therefore analyzed separately.

After the \textsc{JudgmentLens} task, participants additionally rated four system-specific features: explanation depth (FE1), perspective switching (FE2), multimodal presentation (FE3), and AI assistance with source tracing (FE4). After both tasks, they completed a six-item \textsc{JudgmentLens} preference questionnaire and ranked the system modules by usefulness. Exact item wording and response anchors for the SDU, adapted NASA-TLX, \textsc{JudgmentLens}-specific feature, and two-condition preference measures are provided in Appendix~\ref{app:evaluation-materials}, together with the post-task interview core prompts.

\subsubsection{Analysis}

The participant was the unit of analysis and no observations were excluded. We used paired $t$-tests for the four primary outcomes: completion time, rubric-scored comprehension, SDU, and workload. Inspection of the paired-difference distributions did not indicate material departures from normality. We report condition means and standard deviations, the within-participant mean difference ($\Delta=\textsc{JudgmentLens}-\mathrm{PDF}$), 95\% confidence intervals, and Cohen's $d_z$~\cite{lakens2013calculating,cumming2014new}. To control family-wise error across the four primary outcomes, we applied Holm's correction ~\cite{holm1979simple}. Individual feature, preference, and ranking responses are reported descriptively.

\subsubsection{Quantitative Results}

\begin{figure*}[t]
    \centering
    \includegraphics[width=\textwidth]{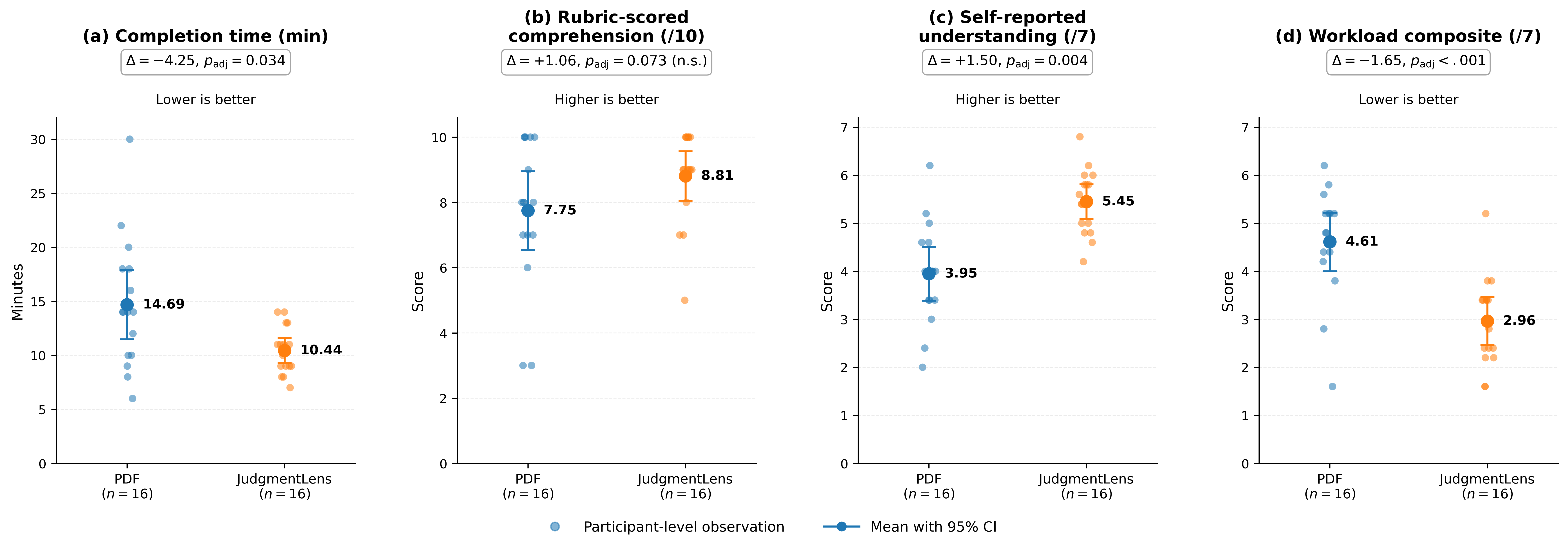}
    \caption{Condition-level distributions for the four primary outcomes
    in the counterbalanced comparison across Cases A and B ($N=16$).
    The four panels show completion time, rubric-scored comprehension,
    self-reported decision understanding, and workload. Points represent
    participant-level observations, while markers and error bars summarize
    the corresponding condition means and 95\% confidence intervals.
    Lower values indicate lower workload. The figure summarizes
    condition-level differences and does not imply that both interfaces
    were applied to the same judgment for each participant.}
    \Description{Four panels compare PDF reading and JudgmentLens on
    completion time, rubric-scored comprehension, self-reported decision
    understanding, and workload. Each panel shows participant observations,
    condition means, and confidence intervals.}
    \label{fig:primary-results}
\end{figure*}

Table~\ref{tab:primary-results} summarizes the controlled comparison. Participants completed the task 4.25 minutes faster with \textsc{JudgmentLens}, a 28.9\% reduction relative to PDF ($t(15)=-2.69$, Holm-adjusted $p=.034$, $d_z=-0.67$). They also reported higher SDU ($t(15)=3.99$, adjusted $p=.004$, $d_z=1.00$) and lower workload ($t(15)=-5.64$, adjusted $p<.001$, $d_z=-1.41$). Figure~\ref{fig:primary-results} visualizes the condition-level
distributions across the four primary outcomes and complements the
summary statistics reported in Table~\ref{tab:primary-results}.

Rubric-scored comprehension was numerically higher with \textsc{JudgmentLens} ($M=8.81$) than with PDF ($M=7.75$), but the difference was not statistically reliable ($t(15)=1.93$, adjusted $p=.073$, $d_z=0.48$). We therefore interpret the controlled evidence as showing improved efficiency and self-reported experience, without claiming an improvement or statistical equivalence in rubric-scored comprehension. The separately analyzed perceived-performance item was lower (better) with \textsc{JudgmentLens} ($M=3.06$, $SD=1.44$) than with PDF ($M=4.13$, $SD=1.41$), but the 95\% CI for the mean difference included zero ($\Delta=-1.06$, 95\% CI $[-2.27,0.14]$).

\begin{table*}[t]
\centering
\caption{Primary counterbalanced within-subject comparison ($N=16$). $\Delta$ is \textsc{JudgmentLens}$-$PDF. Confidence intervals describe paired mean differences; $p_{\mathrm{adj}}$ values use Holm correction across the four primary outcomes.}
\label{tab:primary-results}
\setlength{\tabcolsep}{5pt}
\resizebox{\textwidth}{!}{%
\begin{tabular}{@{}lccccc@{}}
\toprule
\textbf{Measure} & \textbf{PDF $M\pm SD$} & \textbf{JudgmentLens $M\pm SD$} & \textbf{$\Delta$ [95\% CI]} & \textbf{$p_{\mathrm{adj}}$} & \textbf{$d_z$} \\
\midrule
Completion time (min) & $14.69\pm6.03$ & $10.44\pm2.19$ & $-4.25\;[-7.62,-0.88]$ & .034 & $-0.67$ \\
Rubric-scored comprehension (/10) & $7.75\pm2.27$ & $8.81\pm1.42$ & $+1.06\;[-0.11,2.24]$ & .073 & $0.48$ \\
Self-reported understanding (SDU, /7) & $3.95\pm1.05$ & $5.45\pm0.68$ & $+1.50\;[0.70,2.30]$ & .004 & $1.00$ \\
Workload composite (/7) & $4.61\pm1.14$ & $2.96\pm0.94$ & $-1.65\;[-2.27,-1.03]$ & $<.001$ & $-1.41$ \\
\bottomrule
\end{tabular}
}
\end{table*}

Descriptively, the largest SDU item shifts concerned information extraction (PDF: $M=3.88$; \textsc{JudgmentLens}: $M=5.63$), logic mapping ($3.44$ vs. $5.31$), and subsequent inference ($3.88$ vs. $5.56$). Ratings of the four \textsc{JudgmentLens}-specific features were also positive: explanation depth ($M=5.19$, $SD=1.17$), perspective switching ($M=5.50$, $SD=0.97$), multimodal presentation ($M=5.94$, $SD=0.93$), and AI assistance with source tracing ($M=5.50$, $SD=1.26$).

The post-comparison questionnaire showed a similar descriptive pattern. Mean ratings were 5.94 ($SD=1.34$) for overall preference, 5.56 ($SD=1.15$) for perceived efficiency, 6.19 ($SD=1.28$) for reduced cognitive burden, 6.00 ($SD=1.10$) for logical understanding, 5.88 ($SD=1.36$) for confidence, and 5.81 ($SD=1.11$) for willingness to recommend. We use \emph{willingness to recommend} rather than Net Promoter Score because the item used a seven-point scale. In the module ranking, Case Summary received the most first-place selections (11/16), followed by AI Guide (2/16); Case Panorama, Logic Graph, and Evidence each received one. Three rankings were incomplete, so we did not impute missing ranks or calculate a full ranking score.

\subsubsection{How Structure and Source Links Shaped Use}

Participants attributed the reduced effort to persistent organization rather than to text shortening alone. P10 explained, ``With the PDF, there was a lot I had to think through and organize repeatedly. The system laid out the process and logic clearly and saved me a lot of time.'' P1 similarly valued that the system ``summarized the key points, so I did not have to organize and search for them myself,'' while allowing each point to be linked back to the original text. These accounts align with the descriptive SDU item shifts in information extraction and logic mapping: participants began from an external case structure instead of repeatedly reconstructing relationships from linear text.

Source access remained important even when participants preferred the structured views. P16 stated, ``If I really wanted to understand why, I would definitely return to the judgment.'' Thus, source links were experienced as a route for checking an interpretation, not as a reason to treat the generated representation as authoritative. At the same time, the links were not always precise: P6 noted that a source action could open several related passages, requiring them to read through the group again. Other recurring frictions included slow or missing AI responses, dense highlighting, small text, and insufficient onboarding. These limitations show that structural assistance can reduce interpretive work while still introducing interface-learning and navigation costs.

\subsection{Exploratory Contrastive Probe with Conversational AI}

\subsubsection{Motivation, Procedure, and Evidentiary Scope}

The probe examined conversational AI as a different form of mediation, asking what work remains when assistance is organized around user-formulated questions rather than persistent case representations. After the controlled comparison, all participants received Case C and used the original PDF together with DeepSeek. They could interact freely with DeepSeek for 5--8 minutes and could stop early. During the subsequent five-item comprehension test, they could inspect the PDF and conversation history but could not submit new prompts. They then completed the same SDU and workload measures, a retrospective comparison of all three modes, and a second interview. The complete  questions, answer key, item-level scoring rules, assessed dimensions, retrospective three-mode questionnaire, and interview core prompts are provided in Appendix~\ref{app:evaluation-materials}.

This component was not counterbalanced: PDF+DeepSeek always occurred third and always used Case C, whereas PDF and \textsc{JudgmentLens} used Cases A/B in the first two tasks. Interface, case, task order, learning, and fatigue are therefore not identifiable separately. We consequently report no three-condition ANOVA or pairwise significance tests and make no causal claim that \textsc{JudgmentLens} outperformed DeepSeek. Task-level descriptive measures from this probe are reported in the supplementary material solely to document the observed interaction context. Because the probe used a different case, a fixed later position, and a different timing procedure, we do not compare these measures with the controlled conditions.

The retrospective questionnaire is likewise interpreted as participants' comparative perception after experiencing the modes in this fixed sequence. As Table~\ref{tab:retrospective-comparison} shows, participants reported the least cognitive/visual burden and the greatest perceived control with \textsc{JudgmentLens}, while direct PDF access received the highest mean verifiability rating. Forced rankings showed a related distinction: \textsc{JudgmentLens} was most often ranked first for overall preference, source-verification efficiency, and global control, whereas five participants still placed PDF first for consequential decision/trust. P2 provided an incomplete overall-preference ranking, yielding 15 valid responses for that row.

\begin{table*}[t]
\centering
\caption{Descriptive retrospective comparison after all three modes. Ratings used seven-point scales; lower is better only for cognitive/visual burden. Rankings report the number placing each mode first. Because PDF+DeepSeek was administered later with Case C, these values describe perceived experience rather than causal interface effects.}
\label{tab:retrospective-comparison}
\setlength{\tabcolsep}{7pt}
\begin{tabular}{lccc}
\toprule
\textbf{Measure} & \textbf{PDF} & \textbf{PDF+DeepSeek} & \textbf{JudgmentLens} \\
\midrule
\multicolumn{4}{l}{\emph{Retrospective rating, $M\pm SD$}} \\
Cognitive/visual burden $\downarrow$ & $5.69\pm1.25$ & $4.50\pm1.37$ & $2.69\pm1.62$ \\
Verifiability $\uparrow$ & $5.50\pm1.21$ & $3.63\pm1.45$ & $4.63\pm1.54$ \\
Perceived control/agency $\uparrow$ & $2.50\pm1.26$ & $5.06\pm0.85$ & $6.19\pm0.75$ \\
\midrule
\multicolumn{4}{l}{\emph{First-place ranking, $n/N$}} \\
Overall preference & $0/15$ & $3/15$ & $12/15$ \\
Source-verification efficiency & $1/16$ & $2/16$ & $13/16$ \\
Global case control & $1/16$ & $4/16$ & $11/16$ \\
Decision/trust & $5/16$ & $0/16$ & $11/16$ \\
\bottomrule
\end{tabular}
\end{table*}

\subsubsection{What Conversational AI Left to the User}

The interviews revealed an upstream burden that was not captured by answer generation alone. P1 explained, ``If I had not already used the PDF to understand this kind of case, I simply could not have asked these questions. I would have been like a headless fly.'' P15 made the same point while using DeepSeek: ``I did not know what to ask because I did not know which points were important.'' These comments indicate that question answering presupposed an initial model of the case: users first had to decide which issues mattered and turn those issues into prompts.

Participants also remained responsible for coverage, integration, and verification. P3 said that DeepSeek quickly provided a broad outline but might include only what it considered important and omit other details. When returning to the PDF, P3 described the activity as ``examining whether the AI's answer was true, not actively analyzing the case.'' P1 likewise found source checking effortful because a conversational answer did not indicate where each statement appeared in the PDF, sometimes requiring another pass through the document. Participants therefore described conversational AI as useful for local, already-formulated questions, while case-level organization and source checking remained active user work.

We interpret this probe as a contrast between two distributions of human--AI work, not as evidence that one interface universally outperforms the other. Conversational AI externalized answer generation; participants continued to formulate questions, assess coverage, integrate responses, and locate source support. \textsc{JudgmentLens} was designed to externalize more of this upstream sensemaking through persistent, source-linked representations available before, during, and after individual questions.

Taken together, the controlled evidence shows that participants completed judgment-reading tasks faster and reported higher self-reported decision understanding and lower workload with \textsc{JudgmentLens}, while no statistically reliable difference in rubric-scored comprehension was detected. The exploratory probe adds a complementary interaction insight: conversational AI can reduce the cost of answering formulated questions, yet novices may still bear the upstream work of deciding what to ask, organizing answers into a case-level understanding, and verifying those answers against the authoritative judgment.

\section{Discussion}  
\label{sec:discussion}

Our findings suggest that AI-supported judgment interpretation depends on how interpretive work is distributed among the user, the AI system, and the authoritative document~\cite{horvitz1999principles,amershi2019guidelines}. The formative study showed that non-experts spend
substantial effort organizing dispersed facts, connecting evidence to claims,
reconstructing judicial reasoning, and translating conclusions into subsequent
actions. \textsc{JudgmentLens} addresses these demands by externalizing parts
of this work while retaining source access and user control. The controlled comparison provides evidence that the integrated system reduced completion time and subjective workload and increased self-reported decision understanding relative to PDF reading, while no statistically reliable difference in rubric-scored comprehension was detected. The exploratory PDF+DeepSeek probe does not establish a comparative efficiency effect; instead, it provides qualitative insight into how conversational and structurally organized interfaces distribute sensemaking work differently.

\subsection{Redistributing Sensemaking Work between Users and AI}
\label{subsec:discussion_redistribution}

Sensemaking research has long characterized complex information work as a
process of gathering evidence, organizing relationships, constructing
intermediate representations, and iteratively refining interpretations
~\cite{russell1993cost,pirolli2005sensemaking}. From the perspective of
distributed cognition, part of this work can be shifted from internal memory
to external representations that users can inspect and manipulate
~\cite{hutchins1995cognition,zhang1997nature}. Our formative findings suggest
that judicial judgment interpretation contains precisely this kind of hidden
organizational work: readers must determine which facts matter, connect
evidence with disputed propositions, and reconstruct how intermediate
considerations support the final decision.

\textsc{JudgmentLens} makes part of this work explicit through the Case Fact
Core, Timeline, Logic Graph, Evidence view, and coordinated case views.
Importantly, these components do not automate the final act of interpretation.
They externalize orientation, information organization, relationship
construction, and source navigation while leaving evaluation and conclusion
formation with the reader. The strong reductions in completion time and
subjective workload, together with higher perceived information extraction,
logic mapping, and subsequent inference, are consistent with this division of
labor. At the same time, we detected no statistically reliable difference in rubric-scored comprehension between PDF and JudgmentLens. Because a nonsignificant result does not establish equivalence, we cannot conclude that the two interfaces produced the same level of understanding~\cite{lakens2018equivalence}. The controlled evidence most clearly concerns the effort and self-reported experience associated with completing the tasks.

The exploratory PDF+DeepSeek probe offers a qualitative contrast rather than a causal comparison. Because the probe used a different case and always followed the controlled tasks, its task measures cannot distinguish interface effects from case difficulty, learning, order, or fatigue. Participants' accounts nevertheless revealed a recurring difference in how work was distributed. Conversational AI could generate an answer after a question had been formulated, but participants still had to decide what was worth asking, identify possible omissions, integrate multiple responses, and locate supporting passages in the PDF. JudgmentLens was designed to externalize more of this upstream work through persistent case representations available before, during, and after individual questions. The contribution of the probe is therefore not evidence that JudgmentLens outperformed DeepSeek, but an interaction insight: answer generation and case-level sensemaking represent different forms of human--AI collaboration.

This perspective suggests a more precise way to evaluate AI augmentation in
complex knowledge work. The presence of an LLM should not by itself define the
level of augmentation. Instead, researchers should examine \emph{which stages
of sensemaking work are externalized, which remain with the user, and whether
the resulting division lowers the cost of interpretation without obscuring
the underlying evidence}. This framing also explains why the shared Case Fact
Core matters: multiple AI-generated views can remain coordinated projections
of a common case representation rather than becoming independent summaries
that users must reconcile themselves.

\subsection{Making AI Mediation Inspectable through Source Grounding}
\label{subsec:discussion_grounding}

As AI takes on more organizational work, the generated representation itself
becomes part of the user's information environment and therefore a potential
source of opacity. Prior research on explainable AI has shown that explanations
can support understanding, but their usefulness depends on what users need to
accomplish and how explanations fit within the larger interaction
~\cite{miller2019explanation,liao2020questioning}. Research on human-AI reliance
likewise shows that users may over-rely on AI recommendations, even when
explanations are available ~\cite{buccinca2021trust,bansal2021does}. These
concerns are particularly consequential when the underlying document carries
institutional authority.

Participants' accounts suggest that source grounding can function as an interaction mechanism for reducing the practical cost of returning from a system-mediated interpretation to the authoritative document. In \textsc{JudgmentLens}, provenance
links connect summaries, reasoning elements, and AI responses to passages in
the judgment, allowing users to move from an interpretation back to the source
and inspect the surrounding context. This design builds on established
provenance principles, which treat records of origin and transformation as a
basis for judging the reliability and use of information
~\cite{moreau2022provenance}. The contribution is to make provenance operational at
the point of interpretation: rather than attaching a citation as metadata,
the system uses the citation as a navigable path between a mediated
interpretation and its authoritative basis.

The qualitative evidence supports the practical relevance of this mechanism.
Participants repeatedly described the original judgment as necessary for
checking AI-generated explanations, particularly when the interpretation
concerned consequential decisions. In the additional PDF+DeepSeek phase,
participants similarly reported returning to the PDF to compare chatbot
answers with the source. These observations suggest that users did not
necessarily seek an AI answer that could replace the source; they sought an
answer that could be checked against it. This distinction is important because
source grounding should not be framed solely as a strategy for improving
model-side factuality. Retrieval-augmented generation, for example, explicitly
connects generation with retrieved evidence, but the existence of retrieved
documents does not by itself guarantee that users can effectively inspect or
understand the relationship between the output and its evidence
~\cite{lewis2020retrieval}.

We therefore propose \emph{inspectability} as a useful design property of
source-grounded AI for high-stakes documents. A source-grounded system should
make it practically easy for users to answer questions such as: Where did this
claim come from? Which passages support it? Is this statement directly stated
or synthesized? What context might change its interpretation? Evaluating AI
assistance along these dimensions complements conventional evaluation of
generation quality or hallucination rates. It also reframes provenance from a
backend data-management property into a user-facing interaction resource.

\subsection{Supporting User Agency without Substituting Interpretive Authority}
\label{subsec:discussion_agency}

The first two findings imply a broader design tension. AI can substantially
reduce the amount of interpretive work that users must perform, yet reducing
that work can also make it easier to accept an interpretation without engaging
with its basis. This tension is well documented in research on automation bias,
appropriate reliance, and human-AI decision making ~\cite{parasuraman1997humans,buccinca2021trust,he2023knowing}. Human-centered AI therefore emphasizes
systems that increase human capability while preserving meaningful human
oversight and control ~\cite{amershi2019guidelines,shneiderman2022human}.

Our results suggest that preserving agency should be understood operationally
rather than symbolically. Keeping a user ``in the loop'' is insufficient if
the cost of challenging an AI interpretation is too high. In the judicial
setting, practical agency requires that users can bypass an explanation,
return to the source, inspect alternative evidence, question an interpretation,
and ultimately formulate their own conclusion. \textsc{JudgmentLens} supports
these actions by keeping the judgment visible within the same interaction
environment and by coupling generated interpretations with source navigation.

This perspective also clarifies the boundary of our contribution. We do not
claim that \textsc{JudgmentLens} proves that users' epistemic agency is fully
preserved, nor that source grounding eliminates over-reliance. The evaluation did not behaviorally measure epistemic agency as an independent construct, and the retrospective questionnaire captured perceived control rather than error-detection or appropriate-reliance behavior. Moreover, the reviewed study materials did not systematically expose participants to known AI errors. Instead, our findings provide evidence for a design
direction in which the practical cost of exercising agency is reduced. This is
important because users may reasonably want assistance with the burdensome
parts of interpretation while retaining control over consequential judgments.
The design objective is therefore not to minimize AI involvement, but to make
AI involvement compatible with human scrutiny.

This leads to a broader implication for high-stakes AI systems. Rather than
optimizing for outputs that are sufficiently fluent or persuasive that users
no longer need to inspect them, systems should make critical engagement
feasible at the same point where assistance is delivered. In this sense,
source-grounded AI and human agency are complementary: grounding provides the
path for inspection, while agency concerns whether users can meaningfully use
that path to accept, reject, or reinterpret what the system presents.

\subsection{Limitations and Future Directions}
\label{subsec:limitations}

Our study has several limitations.

First, the controlled evaluation involved a relatively small sample of
non-expert readers ($N=16$) and short-term tasks with selected judicial cases.
The results therefore provide evidence about immediate interaction and
comprehension rather than long-term changes in legal behavior or decision
making. Larger and more diverse studies involving self-represented litigants
and users with different levels of digital literacy are needed to examine
whether the observed benefits persist in realistic legal settings.

Second, the primary comparison was conducted between conventional PDF reading
and \textsc{JudgmentLens}. The subsequent PDF+DeepSeek phase served as an exploratory contrastive probe rather than an additional experimental condition. It always followed the controlled tasks and used a separate case; interface, case, task order, learning, and fatigue were therefore confounded. We consequently use the probe to interpret participants' reported interaction experiences rather than as quantitative evidence that JudgmentLens is superior to general-purpose conversational AI. The controlled comparison evaluated JudgmentLens as an integrated system, so effects cannot be attributed to individual components; verification, perceived control, and action-oriented features also remain primarily qualitative and unisolated. Future studies should isolate these mechanisms and include behavioral measures of source checking, error detection, and appropriate reliance.

Third, our rubric-scored comprehension results did not show a reliable
difference between conditions. This limits claims that the system improves factual
knowledge or legal reasoning accuracy. The stronger effects observed for
completion time, subjective workload, and perceived understanding suggest
that future studies should examine efficiency-quality trade-offs more directly,
including delayed comprehension tests and tasks that require users to explain
or justify their interpretations rather than only answer factual questions.

Fourth, source grounding does not guarantee interpretive correctness.
Participants primarily encountered reviewed system outputs in the evaluation,
and we did not systematically inject controlled hallucinations or conflicting
evidence. Future work should expose users to realistic levels of AI error and
measure whether provenance mechanisms actually improve error detection,
appropriate reliance, and correction behavior.

Finally, the present work focuses on judgments in the Chinese legal
context. Although the underlying design principles may be relevant to other
authoritative documents, their applicability to medical records, government
decisions, financial disclosures, or other domains remains an empirical
question. Future research should investigate how the distribution of
sensemaking work and the requirements for source verification change across
different institutional settings.

Together, these limitations point to a broader research agenda: understanding
how AI systems can reduce the practical burden of complex information work
while preserving the conditions under which people can inspect, question, and
remain responsible for consequential interpretations.

\section{Conclusion}
\label{sec:conclusion}

Making judgments publicly available does not necessarily make them meaningfully accessible. Our formative study shows that non-experts struggle not only with legal terminology but with reconstructing relationships among facts, evidence, judicial reasoning, and decisions. \textsc{JudgmentLens} addresses this problem by combining persistent case representations with source-grounded AI assistance. In the controlled comparison, \textsc{JudgmentLens} reduced completion time and workload and increased self-reported decision understanding, while rubric-scored comprehension did not differ reliably. Qualitative findings further suggest that persistent structure and direct paths back to the source shape how users distribute sensemaking work between themselves and the system. More broadly, this work contributes design insights for human-centered AI in high-stakes knowledge domains. We argue that future AI systems should move beyond automation-oriented assistance toward interfaces that support human sensemaking, epistemic agency, and accountable interpretation. By positioning AI as a partner that helps people reason with complex information rather than a substitute for human judgment, \textsc{JudgmentLens} provides a design direction for building trustworthy AI systems in domains where understanding matters as much as efficiency.

\section{Acknowledgments of the Use of AI}

We used AI tools to support both the development of \textsc{JudgmentLens} and the research process. The system integrates DeepSeek models for case representation, structured view generation, and case-grounded question answering. AI tools were also used during research to assist with drafting, editing, and figure development. During the evaluation, participants interacted with the system's LLM-based functions, and the reported results reflect their experiences with these AI-assisted features. The authors reviewed all AI-generated materials and take full responsibility for the system, study design, analysis, and final contents of this paper.


\bibliographystyle{ACM-Reference-Format}
\bibliography{references}

\appendix
\section{Formative Study Materials}
\label{app:formative-materials}

This appendix reports the study instruments in English translation. The
questionnaire was administered in Chinese. The interviews were
semi-structured conversations and the prompts below report the core questions.

\subsection{Phase 1 Questionnaire}

The questionnaire contained the following 15 items. Items marked ``select all
that apply'' permitted multiple responses.

\begin{enumerate}
    \item \textbf{Participant name or nickname.} Participants could provide a
    nickname if they did not wish to disclose their name. This field was used
    only for study administration and was not analyzed.

    \item \textbf{Gender.} Male; Female.

    \item \textbf{Age.} 18--25; 26--35; 36--45; 46--55; 56 or above.

    \item \textbf{Highest educational attainment.} Middle school or below;
    High school/technical secondary school; College/undergraduate degree;
    Master's degree or above.

    \item \textbf{Role in the most recent legal proceeding} (select all that
    apply). Plaintiff; Defendant; Appellant/Appellee; Other.

    \item \textbf{Outcome of that proceeding.} Won; Lost; Partially won;
    Settled/withdrawn.

    \item \textbf{Case type.} Civil dispute (e.g., contract, property,
    marriage, or credit); Administrative case; Criminal case (as a victim or
    a defendant's family member); Labor arbitration/labor dispute; Other.

    \item \textbf{Approximate length of the judgment.} Five pages or fewer;
    6--15 pages; 16--30 pages; More than 30 pages.

    \item \textbf{After receiving the judgment, did you read it in full?}
    Yes, from beginning to end; Read only part of it; Hardly read it; Asked
    another person to read it.

    \item \textbf{How difficult was the judgment to understand?} Five-point
    scale from 1 (very easy) to 5 (very difficult).

    \item \textbf{Which parts of the judgment were most difficult to
    understand?} (select all that apply). Description of case facts; Evidence
    assessment (why evidence was accepted or rejected); Statutory citations;
    The court's reasoning; Operative ruling; Other.

    \item \textbf{What did you do when encountering an unfamiliar legal term?}
    (select all that apply). Search online; Ask a lawyer or legally
    knowledgeable friend; Skip it; Consult legal books or statutory
    provisions; Other.

    \item \textbf{Did difficulty understanding the judgment make you feel
    anxious, helpless, or dissatisfied?} Often; Sometimes; No; Not applicable
    (hardly read it).

    \item \textbf{Would you use a tool that turns a judgment into charts,
    timelines, or flow diagrams and provides AI question answering to help
    explain why the court ruled as it did?} Very willing; Somewhat willing;
    Unsure; Somewhat unwilling; Completely unwilling.

    \item \textbf{If willing, what would you expect this tool to do, or what
    problems should it address?} Open response.
\end{enumerate}

\subsection{Phase 2 Interview Core Prompts}

After restating consent, recording, anonymity, the evaluation interviews followed a shared semi-structured guide while allowing open-ended discussion and condition-specific follow-up probes. The following questions were covered in each session, with their order adapted to the flow of the conversation. The core questions were:

\begin{enumerate}
    \item Please briefly describe the case, your role, its outcome, and the
    issues that mattered most to you.
    \item How did you receive and read the judgment? Which parts did you read
    first, revisit, or skip?
    \item What was difficult to understand (e.g., terminology, statutory
    citations, evidence assessment, the court's reasoning, or the ruling)?
    \item After reading, did you understand why the court reached its result
    and why particular evidence or arguments were accepted or rejected?
    \item What did you do when you did not understand something? Did you ask
    another person, search online, consult legal materials, or use an AI tool?
    \item If you used AI, what did you ask, what was useful or insufficient,
    and how did you judge whether its answer was trustworthy?
    \item How would timelines, diagrams, plain-language explanations,
    case-grounded questions, or links to source passages affect your reading?
    \item What would you want to do after understanding the judgment, and what
    concerns would you have about such a tool (e.g., accuracy, privacy,
    complexity, or over-reliance)?
\end{enumerate}

\subsection{Condensed Analytical Framework}

Table~\ref{tab:formative-analytic-framework} records the analytical bridge
from recurring questionnaire and interview evidence to the four reported
breakdowns. The framework was developed by iteratively
consolidating the initial codes in the study codebook into higher-level
themes, preserving the analytical progression from the original coding
records to the reported findings.

\begin{table*}[t]
\centering
\caption{Condensed analytical framework connecting formative evidence to the
reported sensemaking breakdowns.}
\label{tab:formative-analytic-framework}
\small
\begin{tabular}{@{}p{0.15\textwidth}p{0.34\textwidth}p{0.42\textwidth}@{}}
\toprule
\textbf{Breakdown} & \textbf{Recurring evidence} & \textbf{Analytical interpretation} \\
\midrule
Structural & Difficulty reconstructing chronology and connecting parties,
facts, evidence, rules, and outcomes across long documents. & Readers had to
mentally assemble relationships distributed across linear text. \\
Interpretive & Difficulty with evidence assessment, statutory references, and
the court's reasoning; desire to know why the court ruled as it did. &
Definitions alone did not explain how particular evidence and rules supported
a conclusion. \\
Verification & Reliance on online and AI assistance accompanied by concern
about fabricated or untraceable information. & Assistance needed inspectable
links to the authoritative judgment and visible limits on support. \\
Action & Uncertainty about appeal, obligations, required documents, and what
to do next after receiving a ruling. & Understanding the decision did not
automatically translate into situated procedural preparation. \\
\bottomrule
\end{tabular}
\end{table*}

\section{Evaluation Materials}
\label{app:evaluation-materials}

All evaluation instruments were administered in Chinese; the wording below
is an English translation. Cases A and B were used in the counterbalanced
PDF--\textsc{JudgmentLens} comparison. Case C was used only in the subsequent
exploratory PDF+DeepSeek probe. For every comprehension item, 0 denotes an
incorrect or unsupported response, 1 a response containing one required
component or an otherwise incomplete answer, and 2 a response containing all
required components. The item-specific allocations below operationalized this
rule. Each five-item test had a maximum score of 10.

\subsection{Case A Comprehension Test}

\begin{enumerate}
    \item \textbf{Outcome and sentencing.} What sentences and fines were
    imposed on defendants A and X?\\
    \emph{Answer key:} Each defendant received one year of imprisonment and a
    RMB~1,000 fine.\\
    \emph{Scoring:} 1 point for each defendant when both imprisonment term and
    fine are correct.\\
    \emph{Dimension:} operative outcome.

    \item \textbf{Disputed issue.} Defendant X argued that he had voluntarily
    surrendered. Did the court accept this argument, and why?\\
    \emph{Answer key:} No. Police located and directly apprehended him after
    the co-defendant disclosed his residence; he did not voluntarily submit
    himself, although his later truthful account was treated as a confession.\\
    \emph{Scoring:} 1 point for the rejection; 1 for the absence of voluntary
    surrender/direct apprehension.\\
    \emph{Dimension:} disputed issue and judicial reasoning.

    \item \textbf{Evidence assessment.} Which key evidence did the court rely
    on to determine that the two defendants committed theft jointly?\\
    \emph{Answer key:} Accepted categories included the defendants' accounts,
    victim and witness statements, scene-investigation and identification
    records, recovered physical property, and footprint/appraisal evidence.\\
    \emph{Scoring:} 2 points for at least two valid categories; 1 for one valid
    category.\\
    \emph{Dimension:} evidential support.

    \item \textbf{Procedure and timing.} If a defendant wished to appeal, how
    long after receiving the judgment could the appeal be filed?\\
    \emph{Answer key:} Within 10 days, calculated from the day after receipt.\\
    \emph{Scoring:} 1 point for 10 days; 1 for the correct starting point.\\
    \emph{Dimension:} procedural obligation.

    \item \textbf{Action inference.} Based on the judgment, what should a
    defendant most urgently verify and prepare?\\
    \emph{Answer key:} Verify the date of receipt and resulting appeal
    deadline and decide whether to appeal; if filing a written appeal, prepare
    one original and eight copies. The judgment states that both fines had
    already been paid and that the stolen items had been recovered and
    returned.\\
    \emph{Scoring:} 1 point for receipt date/appeal deadline and decision; 1
    for preparing the written appeal in the stated number of copies.\\
    \emph{Dimension:} subsequent action.
\end{enumerate}

\subsection{Case B Comprehension Test}

\begin{enumerate}
    \item \textbf{Outcome and sentencing.} What principal sentences and fines
    were imposed on L1 (born August 19, 1982), L2, and L3 (born April 8, 1982)?\\
    \emph{Answer key:} L1: one year of imprisonment and RMB~3,000; L2: ten
    months and RMB~3,000; L3: nine months and RMB~3,000.\\
    \emph{Scoring:} 1 point if all imprisonment terms are correct; 1 if all
    fines are correct.\\
    \emph{Dimension:} operative outcome.

    \item \textbf{Disputed issue.} L1 claimed voluntary surrender. Did the
    court accept this argument, and why?\\
    \emph{Answer key:} No. He twice attempted to evade or resist apprehension
    and did not communicate a voluntary submission to police.\\
    \emph{Scoring:} 1 point for the rejection; 1 for evasion/resistance or the
    absence of voluntary submission.\\
    \emph{Dimension:} disputed issue and judicial reasoning.

    \item \textbf{Evidence assessment.} Which key item of evidence did the
    court exclude, and why?\\
    \emph{Answer key:} The Price Appraisal Conclusion. The cable's model,
    source, and parameters were unclear, the identity of the examined object
    was uncertain, and the appraisal procedure lacked a supported reference
    date.\\
    \emph{Scoring:} 1 point for naming the appraisal conclusion; 1 for a valid
    core reason concerning object identity/source or procedural defect.\\
    \emph{Dimension:} evidence acceptance and exclusion.

    \item \textbf{Events and procedure.} Over what period did the adjudicated
    thefts occur, how many incidents were there, and how long was the appeal
    period?\\
    \emph{Answer key:} Four incidents occurred from July 11 to July 30, 2025
    (approximately one month); the July 6 allegation was withdrawn. An appeal
    could be filed within 10 days from the day after receipt.\\
    \emph{Scoring:} 1 point for the period and four incidents; 1 for the
    10-day appeal period.\\
    \emph{Dimension:} temporal reconstruction and procedural obligation.

    \item \textbf{Action inference.} After the judgment took effect, what fine
    and recovery payment did L2 need to verify and arrange?\\
    \emph{Answer key:} RMB~11,400 in total: a RMB~3,000 fine and RMB~8,400 in
    recovered illegal proceeds to be returned to the victim organization.\\
    \emph{Scoring:} 2 points for RMB~11,400; alternatively, 1 point for each
    correct component amount.\\
    \emph{Dimension:} monetary obligation and subsequent action.
\end{enumerate}

\subsection{Case C Comprehension Test}

\begin{enumerate}
    \item \textbf{Outcome and sentencing.} What principal sentences and fines
    were imposed on R and L?\\
    \emph{Answer key:} R received five years and three months of imprisonment
    and a RMB~5,000 fine. L received one year and seven months of imprisonment,
    suspended for three years, and a RMB~3,000 fine.\\
    \emph{Scoring:} 1 point for each defendant when the term, suspension (when
    applicable), and fine are correct.\\
    \emph{Dimension:} operative outcome.

    \item \textbf{Disputed amounts.} How did the court treat the alleged 50
    jin of dried mutton and RMB~4,000 cash attributed to R, and why did it not
    adopt the prosecution's amounts?\\
    \emph{Answer key:} The cash was excluded because it lacked supporting
    evidence. The claimed weight of the dried mutton was uncorroborated, so
    the RMB~1,150 appraisal was not adopted; the court used the RMB~400 resale
    proceeds instead.\\
    \emph{Scoring:} 1 point for the cash and evidential reason; 1 for the meat,
    rejected appraisal, and RMB~400 basis.\\
    \emph{Dimension:} disputed fact and amount determination.

    \item \textbf{Role and evidence.} What core fact supported treating L as
    an accomplice with a secondary role, and which two theft episodes did the
    defendants commit together?\\
    \emph{Answer key:} L stood lookout outside during both jointly committed
    residential thefts. One episode involved a folding bed, dried beef, and a
    whole cow foreleg; the other involved two men's and one women's sheepskin
    robes.\\
    \emph{Scoring:} 1 point for the lookout/secondary-role fact; 1 for both
    joint episodes.\\
    \emph{Dimension:} evidence and principal/accessory role.

    \item \textbf{Precise source location.} After stealing an Agricultural
    Bank card and withdrawing RMB~3,300, how did R obtain the password? Locate
    this fact by PDF page and paragraph.\\
    \emph{Answer key:} A notebook in the stolen bag contained the bank card
    and slips of paper. R saw numbers on one slip, assumed they were the
    password, and gave the card and slip to a stranger who helped withdraw the
    money. This appears on PDF page~2 in the final complete paragraph (the
    paragraph beginning with the January 29, 2017 theft).\\
    \emph{Scoring:} 1 point for the password/stranger account; 1 for the
    correct source location.\\
    \emph{Dimension:} source verification and precise location.

    \item \textbf{Action inference.} If R wished to appeal, what was the
    deadline, and how many originals and copies of the written appeal were
    required?\\
    \emph{Answer key:} Within 10 days from the day after receipt; one original
    and two copies.\\
    \emph{Scoring:} 1 point for the deadline and starting point; 1 for the
    copy counts.\\
    \emph{Dimension:} subsequent procedural action.
\end{enumerate}

\subsection{Post-task Scales}

Unless noted otherwise, items used seven-point response scales. The five
self-reported decision understanding (SDU) items used anchors from 1
(strongly disagree) to 7 (strongly agree); their arithmetic mean formed the
SDU composite.

\begin{enumerate}
    \item I could easily and smoothly extract the judgment's core adjudicative
    facts and outcome.
    \item This presentation helped me quickly clarify how responsibility and
    principal/accessory roles were allocated among multiple defendants.
    \item I could clearly understand which evidence the court accepted, which
    it rejected, and why.
    \item Based on what I read, I could accurately infer the defendants'
    appeal rights and subsequent obligations.
    \item I was confident in my answers to the comprehension test.
\end{enumerate}

The adapted NASA-TLX items used anchors from 1 (very low) to 7 (very high),
except perceived performance, for which 1 indicated very high and 7 very low
performance.

\begin{enumerate}
    \item \textbf{Mental demand:} How much thinking and memory did the judgment
    analysis and question-answering task require?
    \item \textbf{Temporal demand:} How much time pressure did you experience
    while locating answers and understanding the judgment?
    \item \textbf{Perceived performance:} How accurate was your understanding
    of the case and how well did you complete the questions?
    \item \textbf{Effort:} How much effort did you expend to achieve your level
    of understanding and task performance?
    \item \textbf{Frustration:} How much frustration, irritation, or
    helplessness did you experience while reading and organizing the case?
    \item \textbf{Visual and interaction strain:} How burdensome were visual
    fatigue and repeated page turning or scrolling while locating information?
\end{enumerate}

Mental demand, temporal demand, effort, frustration, and visual/interaction
strain were averaged as the workload composite. Perceived performance was
analyzed separately because its anchors ran in the opposite direction.

After the \textsc{JudgmentLens} task, participants rated four feature items
from 1 (strongly disagree) to 7 (strongly agree):

\begin{enumerate}
    \item \textbf{FE1---Explanation depth:} The Beginner, Standard, and
    Professional modes met my information needs at different levels of depth.
    \item \textbf{FE2---Perspective switching:} The Global, Prosecution, and
    Defense perspectives helped me examine the case from multiple viewpoints.
    \item \textbf{FE3---Multimodal presentation:} Compared with long text, the
    timeline, logic graph, and evidence matrix helped me locate key information
    more quickly.
    \item \textbf{FE4---AI assistance:} The AI Guide's case-specific follow-up
    questions and source tracing effectively addressed my questions.
\end{enumerate}

\subsection{Preference Questionnaires}

After completing both controlled tasks, participants rated the following
statements from 1 (strongly disagree) to 7 (strongly agree):

\begin{enumerate}
    \item Compared with conventional PDF reading, I would prefer to use
    \textsc{JudgmentLens} to read and understand judgments.
    \item \textsc{JudgmentLens} helped me locate the core disputed issues and
    ruling more quickly.
    \item Reading with \textsc{JudgmentLens} required less mental effort and
    caused less fatigue than reading the PDF.
    \item The structured views (e.g., timeline and logic graph) reduced the
    difficulty of organizing the case logic.
    \item I was more confident in the accuracy of my comprehension-test answers
    after using \textsc{JudgmentLens}.
    \item I would recommend \textsc{JudgmentLens} to a colleague or friend who
    needed to study a judgment.
\end{enumerate}

Participants then ranked the following modules by usefulness: Case Summary,
Timeline, Logic Graph, Case Panorama, Evidence, Next Actions, Similar Cases,
AI Guide, and Judgment Text.

After the exploratory PDF+DeepSeek probe, participants retrospectively
compared three modes: PDF only, PDF+DeepSeek, and \textsc{JudgmentLens}. For
each of four dimensions they assigned ranks 1 (best), 2, and 3 (worst), using
each rank once: (1) overall preference for everyday use with a complex
judgment; (2) efficiency and precision of locating supporting source text;
(3) ability to establish a global account without becoming lost in details;
and (4) confidence in reaching a conclusion when errors or hallucinations
might occur. They also rated each mode from 1 (very low) to 7 (very high) on:
(5) cognitive load, retrieval friction, and visual fatigue; (6) ease of
checking and verifying a questionable fact; and (7) the sense of directing
the analysis rather than being led by an AI response. Because the probe always
occurred last and used a different case, these retrospective comparisons were
treated descriptively rather than as controlled outcome measures.

\subsection{Post-task Interview Core Prompts}

The evaluation interviews were also semi-structured conversations without a
fixed verbatim guide. The following lists reconstruct the core prompts from
the recordings and transcripts.

\paragraph{After the controlled PDF--\textsc{JudgmentLens} comparison.}
Participants were asked: How did the two reading experiences differ? Which
parts of \textsc{JudgmentLens} were useful or unhelpful, and why? Did the
structured views affect how they found key information or understood case
logic? Did source links affect confidence or verification? Which modules did
they use or avoid? What interaction problems, inaccuracies, or missing
functions did they encounter? What should be improved?

\paragraph{After the exploratory PDF+DeepSeek probe.}
Participants were asked: How did PDF+DeepSeek differ from PDF alone and from
\textsc{JudgmentLens}? How did they decide what to ask, and were important
questions difficult to formulate? Did they trust answers directly or compare
them with the PDF? How did conversational text compare with timelines,
graphs, and persistent case representations? Which mode best supported a
global account, source checking, and a sense of control? What limitations or
improvements did they identify?

\section{Additional Implementation Details}
\label{app:implementation}

This appendix provides implementation details necessary to interpret the
evaluated prototype and its source-grounding mechanisms.

\subsection{Deployment and Processing Pipeline}

\textsc{JudgmentLens} was implemented as a web application with a
React/TypeScript frontend and FastAPI backend. Asynchronous preparation jobs
were coordinated through Redis and Celery, while SQLite and local file storage
managed application records, generated case artifacts, and uploaded documents.
The evaluated prototype ran on a study-controlled cloud instance, and configured
externally hosted model endpoints were used for generation.

The upload pipeline accepted PDF, DOCX, and legacy DOC files. Parsers extracted
document text, section structure, paragraph boundaries, case metadata, cited
statutes, monetary amounts, and procedural deadlines. The system maintained
two aligned representations: an \emph{Original Representation} containing the
extracted judgment text and a \emph{Redacted Representation} in which
rule-based detectors masked common identifying fields. Redaction reduced
disclosure but was not treated as complete de-identification. The selected
privacy mode determined which representation was passed to downstream
generation.

\subsection{Case Representation and Generation}

After extraction, the backend constructed a versioned \emph{Case Fact Core}
containing parties, claims or charges, established facts, evidence, judicial
reasoning, rulings, deadlines, and monetary or sentencing obligations. Each
reviewable item retained source-anchor identifiers. The Summary, Timeline,
Case Panorama, Logic Graph, Evidence, Next Actions, Similar Cases, and
question-starter modules were generated as coordinated projections of this
shared representation rather than as independent summaries.

Perspective and explanation depth changed ordering, emphasis, vocabulary, and
visible detail while preserving the same underlying case structure. Generation
outputs were schema-validated and checked for source alignment. When the
preferred generation path failed, deterministic fallbacks produced a more
limited representation from extracted facts and source text. Versioned
prompts, schemas, and content hashes were used to identify cached projections
and invalidate stale outputs when source-derived content changed.

The evaluated system used task-aware routing across configured DeepSeek
endpoints. The more capable endpoint was used for Fact Core construction and
Professional-depth generation, while routine projections and interactive
responses were routed to a lower-latency endpoint. General judgment retrieval
used a locally executed Chinese text encoder, while a separate encoder was
used for retrieval over the curated classic-case collection.

\subsection{Source Support and AI Guide}

Each source anchor stored a paragraph or chunk identifier, a supporting
excerpt, and information required to relocate the passage. Generated items
were assigned qualitative source-support states such as \emph{Direct Source},
\emph{Source-supported Synthesis}, \emph{Weak Source}, or
\emph{Requires Review}. These states were surfaced to users as support cues;
they were not calibrated probabilities or estimates of legal correctness.

The AI Guide retrieved paragraph-level passages from the active judgment before
generation. Retrieved passages supplied the immediate documentary context and
source references, while server-sent events streamed responses to the
interface. Selecting a source-linked item updated a shared source-focus state,
opened the Judgment Text panel, positioned the document at the corresponding
passage, and highlighted the relevant text while preserving the reader's
current view and settings.

Schema validation, source-alignment checks, provider-specific limits, caching,
and deterministic fallbacks were used to isolate malformed or unavailable
modules. These mechanisms supported graceful degradation and source
inspection, but did not guarantee correctness of generated legal
interpretations.

\subsection{Frozen Evaluation Materials}

Before evaluation, we froze the prompts, schemas, and model-routing
configuration. For Cases~A and~B, we pre-generated the combinations of three
reading perspectives, three explanation depths, and two privacy modes,
yielding 18 Reading Packs per case. Researchers manually checked
participant-facing facts, party roles, disputed issues, rulings, dates,
monetary or sentencing obligations, and linked source passages against the
corresponding judgments.

All frozen packs passed this study-specific review without a material factual
or source-alignment discrepancy. This procedure stabilized the materials used
in the evaluation but does not establish general model accuracy. Prompt and
schema versions, deployment instructions, and the review checklist are
included in the anonymized artifact.

\end{document}